\RequirePackage{fix-cm}
\documentclass[smallextended]{svjour3}       
\smartqed  
\usepackage{graphicx}
\usepackage[utf8]{inputenc} 
\usepackage[T1]{fontenc}    
\usepackage{hyperref}       
\usepackage{url}            
\usepackage{booktabs}       
\usepackage{amsfonts}       
\usepackage{nicefrac}       
\usepackage{microtype}      
\usepackage{lipsum}
\usepackage{graphicx}
\usepackage{comment}
\usepackage{multirow}
\usepackage{subcaption}
\usepackage{amsmath}

\usepackage[square,numbers]{natbib}
\begin{document}

\title{NodeSim: Node Similarity based Network Embedding for Diverse Link Prediction
}


\author{Akrati Saxena         \and
        George Fletcher \and
        Mykola  Pechenizkiy 
}


\institute{Akrati Saxena, George Fletcher, Mykola  Pechenizkiy \at
              Department of Mathematics and Computer Science\\
  Eindhoven University of Technology, Netherlands \\
              \email{a.saxena@tue.nl,}   {g.h.l.fletcher@tue.nl}, {m.pechenizkiy@tue.nl}       
}


\date{Received: date / Accepted: date}

\maketitle

\begin{abstract}
In real-world complex networks, understanding the dynamics of their evolution has been of great interest to the scientific community. Predicting future links is an essential task of social network analysis as the addition or removal of the links over time leads to the network evolution. In a network, links can be categorized as intra-community links if both end nodes of the link belong to the same community, otherwise inter-community links. The existing link-prediction methods have mainly focused on achieving high accuracy for intra-community link prediction. In this work, we propose a network embedding method, called NodeSim, which captures both similarities between the nodes and the community structure while learning the low-dimensional representation of the network. The embedding is learned using the proposed NodeSim random walk, which efficiently explores the diverse neighborhood while keeping the more similar nodes closer in the context of the node. We verify the efficacy of the proposed embedding method over state-of-the-art methods using diverse link prediction. We propose a machine learning model for link prediction that considers both the nodes' embedding and their community information to predict the link between two given nodes. Extensive experimental results on several real-world networks demonstrate the effectiveness of the proposed framework for both inter and intra-community link prediction. 

\end{abstract}

\section{Introduction}

In online social networks (OSNs), nodes are organized into communities, where a community represents a group of nodes having similar characteristics, such as similar interests, opinions, or beliefs \citep{clauset2004finding}. The links between the nodes belonging to the same community are referred to as \textit{intra community links}, and the links between the nodes belonging to different communities are referred to as \textit{inter community links}. In social networks, intra-community links are driven by the effect of homophily \citep{mcpherson2001birds} as similar nodes prefer to connect with each other. The formation of inter-community links is still not well explored in the literature; however, it can be explained by different complex phenomena, such as triadic closure and weak ties \citep{granovetter1983strength}. In real-world networks, it is observed that the number of intra-community links is more than the number of inter-community links \citep{saxena2016evolving}. The evolution of social networks is regulated by the formation of new links in the network.

In OSNs, we recommend more probable, but not existing links as promising connections to help users in making new friends, and a user having more friends will be more loyal towards the website \citep{benevenuto2009characterizing,wilson2009user}. However, forming the right kind of links is very important as the opinion of a user is highly influenced by the opinion of its neighbors \citep{saxena2020mitigating}. In the recent era, scientists have focused on increasing the diversity in the network so that the users receive information on a topic from different viewpoints before making their opinion \citep{masrour2020bursting}. It is very crucial that a user receives the information from other users having different perspectives to mitigate the negative impact of fake propaganda, false information, or fake news spreading on the network \citep{aslay2018maximizing}. Hence, it is required that a user has a \textit{diverse} neighborhood by having connections with different communities. In social networks, more inter-community links should be promoted to increase diversity. The link recommendation system plays an important role in forming new links and transforming the network evolution.

Initially, researchers proposed link prediction methods based on the similarity of the nodes \citep{zhou2009predicting}. These methods compute the similarity of a pair of nodes based on network structure, and more similar nodes are more likely to form a link. These methods are also often referred to as \textit{classic} or \textit{heuristic link prediction methods}. The well known classic method includes Jaccard coefficient \citep{liben2007link}, Adamic Adar index \citep{adamic2003friends}, resource allocation index \citep{zhou2009predicting}, preferential attachment index \citep{barabasi1999emergence}, and so on. These methods were extended to include community structure to improve the link prediction accuracy; however, most of the methods improved the total accuracy by improving intra-community link prediction accuracy \citep{valverde2013exploiting,jeon2017community}. 

In recent works, network characteristics have been studied using network embedding where the network is represented in a low dimensional latent space \citep{grover2016node2vec}. In network embedding techniques, the aim is to embed similar nodes closer to each other. Most of the existing network embedding methods focus on embedding the nodes closely if they belong to the same community and therefore have high accuracy for the node classification task and intra-community link prediction. 

In our work, we propose a network embedding method, called NodeSim embedding, which considers both the nodes' similarity and their community information while generating the network embedding. In the learned embedding, the nodes belonging to the same community will be embedded closely, and the nodes belonging to different communities will be embedded closer based on their neighborhood similarity. Therefore, the generated embedding preserves the structural properties of the network and is efficient in predicting diverse promising links. Next, we propose a link prediction method that trains a logistic regression model using node pair embedding and their community information to predict both the inter and intra-community links with high accuracy. This is the first work that uses community information for learning the link prediction model and achieves higher accuracy for both types of links. The experiments are performed to show the accuracy and efficiency of the proposed method on real-world networks. The results show that the proposed method outperforms the state-of-the-art methods on all the datasets.

The paper is structured as follows. In Section \ref{relatedwork}, we discuss the state of the art literature on link prediction by focusing on network embedding techniques. In Section \ref{proposemethod}, we discuss the proposed methods, including (i) NodeSim network embedding method and (ii) link prediction method. In Section \ref{results}, we discuss experimental results on real-world networks, including the performance, sensitivity, scalability, and robustness analysis of the proposed method. The paper is concluded in Section \ref{conclusion} with future directions. 

\section{Related Work}\label{relatedwork}

Link prediction is a very well-known problem in network science and has been applied to predict missing links in different types of networks, such as friendship networks, collaboration network, and chemical networks. Initially, researchers proposed heuristic methods, also known as classic methods, which consider the similarity of two nodes to predict the link between them. The well known classic methods include Jaccard Coefficient, Adamic Adar Index, resource allocation index, preferential attachment index, and so on. The initially proposed methods only considered the neighborhood information of the nodes for link prediction and did not consider the network topology. Then, researchers extended these methods that also considered the network structure properties like community structure to predict the links \citep{jeon2017community,valverde2012structural,valverde2013exploiting}. However, most of these methods improved the overall accuracy of link prediction by improving the accuracy of intra-community link prediction. The main benefit of using classic methods is that these methods do not need any training and are comparatively faster. 

Another class of link prediction methods uses machine learning models, such as probabilistic graphical models \citep{clauset2008hierarchical,wang2007local}, matrix factorization \citep{scripps2008matrix,menon2011link}, supervised learning methods \citep{al2006link,lu2010supervised,benchettara2010supervised}, and semi-supervised learning methods \citep{kashima2009link,hu2017lpi}. These machine learning methods provide good accuracy though they suffer from the class imbalance problem as the number of existing links in a network are significantly fewer than the number of non-existing links.

In recent years, network embedding techniques have been used to study networks and to propose solutions for various network analysis problems. The network embedding methods can be categorized into three categories based on the structural proximity considered while generating the embedding, (i) microscopic structure embedding, which considers local proximity of nodes, such as first-order \citep{tang2015line,wang2016structural}, second-order \cite{tang2015line} or high-order proximity \citep{perozzi2014deepwalk,grover2016node2vec,cao2015grarep}, (ii) mesoscopic structure embedding, which captures hierarchical and community structural proximity \citep{du2018galaxy,keikha2018community,li2019learning}, and (iii) network properties preserved embedding, which captures global network properties, such as network transitivity or structural balance \citep{ou2016asymmetric,ou2015non}.


In the existing mesoscopic network embedding, the main focus has been either on the hierarchical embedding where the users belonging to the same hierarchy should be embedded together \citep{du2018galaxy} or on the intra-community proximity where the nodes belonging to one community should be embedded closely \citep{keikha2018community,li2019learning}. In hierarchical or structural role proximity, the nodes playing the same roles are embedded closely; for example, the nodes having a similar degree or similar influential power should be embedded closer \citep{lyu2017enhancing}. In this work, we propose the NodeSim network embedding method, which considers both (i) high-order proximity by the similarity of the nodes and (ii) mesoscopic structure by the network communities while generating the embedding. In NodeSim embedding, the nodes belonging to one community are clustered together, and the similar nodes belonging to different communities are embedded closer. The proposed embedding captures a richer diverse neighborhood of the nodes that is further verified using the link prediction. 



\section{The Proposed Method}\label{proposemethod}

In this section, we first discuss the required network properties for our work. Next, we discuss our proposed NodeSim embedding method to learn the feature representation of the nodes and the proposed link prediction method.

\subsection{Community Structure}

In real-world complex networks, nodes connect with each other if they have similar properties. A group of nodes that are densely connected with each other is referred to as a community \citep{newman2006modularity}. The community label of a node $u$ is denoted by $C_u$. If both end nodes of a link $(u,v)$ belong to the same community, it is referred to as an intra-community link, and $C_{(u,v)} =1$ for an intra-community link. If both end nodes belong to different communities, then the link $(u,v)$ is referred as an inter-community link and $C_{(u,v)} =0$. 

In most real-world networks, the ground truth community information is not available. In literature, several community detection methods have been proposed to identify communities using network structure if the ground truth information is not known. In this work, we apply the highly used community detection method, known as the Louvain method \citep{blondel2008fast}, to identify the communities if the ground truth information is not known. 

\subsubsection{Louvain Community Detection Method}
Louvain method \citep{blondel2008fast} uses two-step greedy optimization to optimize the modularity of a community partition of the network. First, the method optimizes the modularity locally to find small communities. In the second step, it merges all nodes belonging to the same community and creates an aggregated network where each node represents a community. These steps are performed iteratively until we achieve the maximum modularity and the obtained communities are returned. 

\subsection{Node-Pair Similarity} 

In a network, two nodes connect with each other if they have some common interest or characteristics, and therefore, a link between a pair of nodes is the first indication that they are similar. However, these binary/unweighted connections cannot capture the complete information of the system as each connection is not equally important. A better way of representing the network is with weighted edges, where edge-weight denotes the strength of the connection \citep{saxena2020survey}. For example, in a friendship network, the weight of an edge can be computed based on the intimacy of the relationship or frequency of the communication \citep{onnela2007analysis}. The similarity of a node pair $(u,v)$ is denoted as $Sim(u,v)$. 

In most real-world networks, the edge-weight data is not available as it is not feasible to collect all the required information for computing the strength of each connection. In network science, there have been proposed methods to compute the similarity of a node-pair based on their neighborhood connectivity in the network structure. Some of the well-known methods are the number of common neighbors \citep{newman2001clustering}, Jaccard coefficient \citep{liben2007link}, Adamic-Adar \citep{adamic2003friends}, resource-allocation \citep{zhou2009predicting}, hub promoted index \citep{ravasz2002hierarchical}, and so on, which compute a node-pair similarity based on their local-neighborhood proximity.


In this work, we will use the Jaccard coefficient to compute a node pair's similarity in unweighted networks. The Jaccard coefficient for a node pair $(u,v)$ is defined as, $JC(u,v)=\frac{|\Gamma(u) \cap  \Gamma(v)|}{|\Gamma(u) \cup  \Gamma(v)|}$, where $\Gamma(u)$ is the set of neighbors of node $u$.

\subsection{NodeSim Network Embedding}


For a given graph $G(V, E)$, the network embedding method learns the mapping $\Phi: V \rightarrow  \mathbb{R}^d$, where $d$ is the dimension of the embedding space. In recent works, the Skip-gram model has been used to generate the network embedding by representing the network as a document where the nodes are corresponding to the words \citep{perozzi2014deepwalk}. In a network, a sampled sequence of nodes is considered the same as an ordered sequence of words in a document. The simplest way to generate the ordered sequence of nodes is by using random walks. 

In the random walk \citep{lovasz1993random}, if the random walker is at node $u$, the probability that the random walker will move to node $v$ is defined as,

\begin{equation*}
    P_{uv}= \left\{\begin{matrix}
1/deg(u), &  if (u,v) \in E \\ 
0, & otherwise
\end{matrix}\right.
\end{equation*}

The random walk method does not consider the network structure properties while sampling the nodes. In recent works, different sampling methodologies have been explored to sample the network to learn feature representations of the network \citep{grover2016node2vec,de2018combining}. 
However, the proposed methods do not consider the meso-scale properties, such as community structure, while exploring the network. In this work, we propose a random walk based sampling method, called NodeSim Random Walk, that captures the neighborhood of the node by considering both the nodes' similarity as well as the meso-scale community structure of the network. 

\subsubsection{NodeSim Random Walk}

In network embedding, the focus is to embed similar nodes closer. The simplest way to capture the node similarity during the random walk would be to bias the edge probability based on the similarity of its end nodes. However, this will ignore the meso-scale property of the network that is captured through the community structure. In NodeSim random walk, the edge-probabilities are assigned based on both the similarity of the nodes and community structure.

In NodeSim Random walk, the unnormalized probability $p_{uv}$ to move from node $u$ to node $v$ is defined as, 

\begin{equation}
p_{uv}=    \left\{\begin{matrix}
\alpha \cdot (Sim(u,v) +1/deg(u)), &  if (u,v) \in E \; and \; C_{(u,v)} = 1 \\ 
\beta \cdot (Sim(u,v) +1/deg(u)), &  if (u,v) \in E \; and \; C_{(u,v)} = 0 \\ 
0, & otherwise
\end{matrix}\right.
\end{equation}

The probabilities are normalized for each node $u$ with respect to all of its neighbors. So, the probability to move from node $u$ to node $v$ is computed as, $P_{uv}= p_{uv} \cdot w_{u}$ where $w_{u}$ is the normalizing factor for node $u$.

In this work, the similarity of the nodes is computed using the Jaccard Coefficient. Figure \label{iicne} explains edge-probabilities for NodeSim random walk, where the network has two communities shown by red and blue nodes, and the edges $(u,v)$ and $(u,w)$ are inter and intra-community edges, respectively, which are labeled with $p_{uv}$ and $p_{uw}$, respectively.

\begin{figure}[]
	\centering	
	\includegraphics[width=\linewidth]{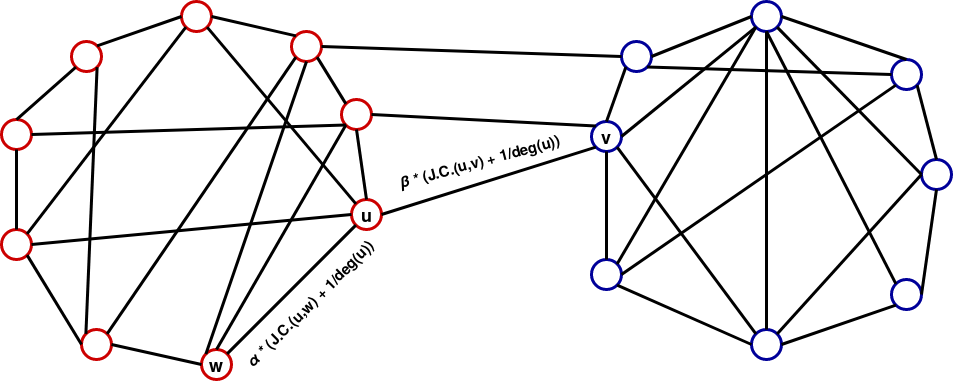}
	\caption{NodeSim Random Walk probabilities for inter and intra community nodes.}
	\label{iicne}
\end{figure}

Intuitively, parameters $\alpha$ and $\beta$ control how the random walker explores the neighborhood. A higher value of $\alpha$ shows that the walker will prefer to sample more similar nodes from the same community, and a higher value of $\beta$ shows that the walker will put a higher weight to explore the inter-community neighborhood of the node. 

\subsubsection{Learn Embedding}

Once the ordered sequences of nodes are generated using NodeSim random walk, the network embedding is learned using the Skip-gram model \citep{mikolov2013efficient}. 
The network embedding method learns a mapping for each node $u \in V$ to a $d$-dimension embedding space that represents the $d$-dimensional feature representation of node $u$ based on its structural role. The network embedding is denoted as $\Phi: u \in V \longrightarrow \mathbb{R}^{|V| \times d}$, where $\Phi$ can be considered a $|V| \times d$ size matrix that is learned by solving a maximal likelihood optimization problem. 

In the skip-gram model, given the corpus, the neighborhood of a word is defined using a sliding window over the consecutive words. In networks, we generate the ordered sequence of nodes using sampling methods. For example, if NodeSim random walker visits the following nodes $\{u_1, u_2, \cdots  u_i, \cdots  u_l\}$, they will be referred to as an ordered sequence of nodes. The neighborhood of a node $u_i$ will be defined by considering $k-1$ nodes visited before and after node $u_i$ during the sampling, where $k$ is the window size or context of the node. For every node $u_i \in V$, $N_{NS}(u_i) \subset V$ denotes the neighborhood of node $u_i$ in the network that is generated through the NodeSim sampling method with the given context $k$.

In the skip-gram model, the network embedding is learned based on the likelihood of a node $u_i$ co-occurring with other neighborhood nodes within the context $k$ in the NodeSim random walk. We, therefore, optimize the following optimization function that aims for maximizing the probability of observing a node in the neighborhood of node $u_i$, given its feature representation $\Phi(u_i)$,

\begin{equation}\label{eq1}
    \underset{\Phi}{maximize} \sum_{u_i \in V} logPr ( N_{NS}(u_i) | \Phi(u_i))
\end{equation}


The optimization problem is solved using two assumptions. The first assumption is conditional independence, that the probability of observing a node in the neighborhood of the source node is independent of observing any other node in its neighborhood given the feature representation of the source node, so, 

\begin{equation}\label{eq2}
    Pr ( N_{NS}(u_i) | \Phi(u_i))= \Pi_{u_j \in N_{NS}(u_i)}Pr(u_j| \Phi(u_i))
\end{equation}


The second assumption is the symmetry that considers the pairwise similarity of a source node and its neighborhood node in the feature space. Therefore, we estimate the probability of a node $u_j$ co-occurring with node $u_i$ using the softmax function,

\begin{equation}\label{eq3}
    Pr (u_j |\Phi(u_i) ) = \frac{exp(\Phi(u_j) \cdot \Phi(u_i))}{\sum_{v \in V} exp(\Phi(v) \cdot \Phi(u_i)) }
\end{equation}

Finally, using both assumptions, the objective function given in Equation \ref{eq1} is computed as, 

\begin{equation}\label{eq4}
    \underset{\Phi}{maximize} \sum_{u_i \in V} \left ( -logZ_{u_i} + \sum_{u_j \in N_{NS}(u_i)}\Phi(u_j) \cdot \Phi(u_i) \right )
\end{equation}

where $Z_{u_i} = \sum_{v \in V} exp(\Phi(u_i) \cdot \Phi(v)$ is expensive for large-scale networks and it is approximated using negative sampling method \citep{mikolov2013distributed}. Equation \ref{eq4} is optimized using SGA (stochastic gradient ascent) over the features $\Phi$ \citep{grover2016node2vec}. 

\subsubsection{Complexity}

The complexity of the proposed network embedding method depends on two major steps, (i) identify the communities and (ii) NodeSim embedding learned using the Skip-gram model. The complexity of the community detection method and Skip-gram model is well defined in the literature, so we briefly discuss the complexity of our method. In our implementation, we have used the Louvain community detection method having complexity $O(n \cdot logn)$. Once the community structure is identified, the complexity to generate the probability distribution for NodeSim random walk is $O(m)$. The complexity for learning embedding using the skip-gram model with negative sampling is $O(nkl \gamma (d+ d log(n)))$, where $d$ denotes the number of dimensions, $l$ denotes the walk length, $k$ denotes the window size, and $\gamma$ denotes the number of random walks. So, the overall complexity is $O(nlogn + m + nkl \gamma (d+ d log(n)))$. 

\subsection{Link-Prediction Method}

The link prediction method first generates the feature representation of given node pairs and then train a logistic regression model using the feature representation of node pairs and their community information.

\subsubsection{Feature Representation of Node Pair}

The feature representation of a pair of node $(u,v)$ is generated by applying a binary operator on the feature representation of node $u$ and $v$. The most common operators are mentioned below.

\begin{enumerate}
    \item Average: $e_i(u,v)=\frac{\Phi_i(u) + \Phi_i(v)}{2}$
    
    \item Weighted-L1: $e_i(u,v)=|\Phi_i(u) - \Phi_i(v)|$
    
    \item Weighted-L2: $e_i(u,v)=|\Phi_i(u) - \Phi_i(v)|^2$
    
    \item Hadamard: $e_i(u,v)=\Phi_i(u) * \Phi_i(v)$
\end{enumerate}

$\Phi_i(u)$ denotes the $i_{th}$ feature of node $u$, and $e_i(u,v)$ denotes the $i_{th}$ feature of a node pair $(u,v)$. In this way, a $d$-dimension feature vector is generated for each node-pair using the $d$-dimension feature representation of the corresponding nodes.

\subsubsection{Link Prediction Model}

For link prediction, a logistic regression model is trained using features of the node-pair and their community information, with the output having the existent/non-existent information of the link between the given node-pair. The input features for a node pair $(u,v)$ is generated as, $f(u,v)= (e(u,v) || C_{(u,v)})$, where $||$ is concatenation operator and $C_{(u,v)}$ is $1$ if both nodes $u$ and $v$ belong to the same community, otherwise $0$. The output parameter is $1$ or $0$ if there exists a link between the given pair of nodes or not, respectively. We have shown results for all four operators applied on $e(u,v)$.

\section{Experimental Analysis}\label{results}

In this section, we discuss baseline methods, datasets, and experimental results.

\subsection{Baseline Methods}

The proposed method is compared with both types of link prediction methods (i) similarity-based heuristic methods and (ii) network embedding based methods.

We compare with the following three heuristic methods based on network structure.

\begin{enumerate}
    \item Jaccard Coefficient (JC) \citep{liben2007link}: $JC(u,v)=\frac{|\Gamma(u) \cap  \Gamma(v)|}{|\Gamma(u) \cup  \Gamma(v)|}$
    \item Adamic Adar (AA) \citep{adamic2003friends}: $AA(u,v)=\sum_{ w \in (\Gamma(u) \cap  \Gamma(v))} \frac{1}{log|\Gamma(w)|}$
    \item Resource Allocation (RA) \citep{zhou2009predicting}: $RA(u,v)=\sum_{ w \in (\Gamma(u) \cap  \Gamma(v))} \frac{1}{|\Gamma(w)|}$
\end{enumerate}

We compare our method with the following network embedding based link-prediction methods.

\begin{enumerate}\setcounter{enumi}{3}
    \item DeepWalk \citep{perozzi2014deepwalk}: Deepwalk method learns the network embedding using the skip-gram model on the ordered sequence of nodes generated using random walk.
    
    \item Node2Vec \citep{grover2016node2vec}: Node2Vec is an extension of DeepWalk where the walker has different probabilities for moving to its neighbors, and the probability to move to the next node depends on its distance from the previously visited node. Once the nodes are sampled, the network embedding is learned using the skip-gram model. We have used the code provided by the authors at \url{https://github.com/aditya-grover/node2vec}.
    
    \item NECS \citep{li2019learning}: Network Embedding with Community Structural information (NECS) uses nonnegative matrix factorization to generate nodes' embedding, which preserves the high-order proximity. The final network embedding is learned by jointly optimizing the consensus relationship between the nodes' representation and the community structure. We have used the implementation provided by the authors at \url{https://github.com/liyu1990/necs}. %
    
   For DeepWalk, Node2Vec, and NECS methods, the node-pair embedding is generated using the Hadamard operator, and then the logistic regression model is trained for the link prediction as mentioned in these works.

    \item Splitter \citep{epasto2019single}: This network embedding method learns multiple embedding of each node based on the principled decomposition of the ego-network. These multiple representations of a node denote its embedding with respect to the local communities it belongs to. The implementation is provided by the authors at \url{https://github.com/google-research/google-research/tree/master/graph_embedding/persona}. For link prediction, we used the method discussed in their paper. For each node pair $(u,v)$, the similarity score is computed using the dot product of their embedding. In the persona graph, each node has multiple embedding, so we compute the similarity score for each combination of their embedding, and the maximum score is returned as the final similarity score. Once the final similarity score is computed, the link prediction is performed using the same method as for similarity-based heuristic methods.
\end{enumerate}

\subsection{Datasets}\label{datasets}

We perform experiments on real-world networks, and their details are mentioned in Table \ref{datasets}. Facebook is a small subgraph extracted from Facebook social networking website. GrQc, Hep-th, Hep-ph, and Astro-ph are co-authorship networks extracted from ArXiv for General Relativity and Quantum Cosmology, High Energy Physics Theory, High Energy Physics Phenomenology, and Astro Physics research areas, respectively. In all the networks, the communities are detected using the Louvain Method, and a community label is assigned to each node based on which community it belongs to. A node pair is referred to as intra-community node pair if both the nodes belong to the same community; otherwise, it will be referred to as inter-community node pair.

\begin{table}[htbp]
\centering
\caption{Datasets}
\label{datasets}
\resizebox{\textwidth}{!}{
\begin{tabular}{|l|r|r|r|r|}
\hline
\textbf{Network}         & \textbf{\#nodes} & \textbf{\#edges} & \textbf{\#communities} & \textbf{Ref} \\ \hline
\textbf{Facebook}        & 4039             & 88234            & 16            &  \citep{leskovec2012learning}   \\ \hline
\textbf{GrQc}            & 4158             & 13422            & 42            &   \citep{leskovec2007graph}   \\ \hline
\textbf{Hep-th}          & 8638             & 24806            & 50            &  \citep{leskovec2007graph}    \\ \hline
\textbf{Hep-ph}          & 11204            & 117619           & 38            &  \citep{leskovec2007graph}    \\ \hline
\textbf{Astro-ph}        & 17903            & 196972           & 37            &  \citep{leskovec2007graph}   \\ \hline
\end{tabular}
}
\end{table}

To generate the training and testing data, we follow the same methodology as used in \citep{epasto2019single,grover2016node2vec}; however, we maintain the ratio of inter and intra-community links that is not considered in previous studies. First, we remove $10\%$ of inter-community and $10\%$ of intra-community edges from $E$ uniformly at random and put them in set $E_{lp}$ that will be used for link prediction. While removing the $10\%$ edges, it is ensured that the network remains connected. The remaining $90\%$ edges are referred to as $E_{ne}$, and $G(E_{ne}, V)$ will be used to generate network embedding. 

For link prediction task, the same number of inter and intra-community node pairs for non-existent links are chosen uniformly at random, as we have in $E_{lp}$. These sampled links will work as negative cases and are added to set $E_{lp}$. If a link is formed between a given node pair, then it is referred to as a positive case; otherwise, it will be referred to as a negative case. To create train and test data, the node pairs in $E_{lp}$ are split into $E_{train}$ and $E_{test}$, and while splitting, we ensure that the ratio of intra and inter-community node pairs is maintained for both positive and negative cases. The default train and test ratio is $(.5:.5)$ if it is not mentioned explicitly. In heuristic and Splitter link Prediction methods, a node pair is predicted positive if the similarity score for this pair is higher than the similarity score of $25\%$ positive train cases. 

\subsection{Performance Study}

First, we compare NodeSim method with baselines, and ROC-AUC value is computed for all test cases, intra-community and inter-community test cases as shown in Table \ref{rocauc}. The table shows the best results observed for different parameter settings used in different methods, and each experiment is repeated five times to compute the average. The dimension of network embedding is $d=128$. The results show that the proposed NodeSim method with Hadamard operator for node pair embedding outperforms all baseline methods for both types of links. The NECS method for the Astro-ph network was not completed in 48 hours on the server, so the values are not mentioned. 

\begin{table}[]
\caption{ROC-AUC for link prediction}
\label{rocauc}
\resizebox{\textwidth}{!}{
\begin{tabular}{|l|l|r|r|r|r|r|}
\hline
\textbf{Method\textbackslash{}Datasets}             &                & \multicolumn{1}{l|}{\textbf{Facebook}} & \multicolumn{1}{l|}{\textbf{GrQc}} & \multicolumn{1}{l|}{\textbf{Hep-th}} & \multicolumn{1}{l|}{\textbf{Hep-ph}} & \multicolumn{1}{l|}{\textbf{Astro-ph}} \\ \hline
                                                    & \textbf{Total} & 0.713                                  & 0.756                              & 0.735                                & 0.741                                & 0.750                                   \\ \cline{2-7} 
                                                    & \textbf{Intra} & 0.721                                  & 0.788                              & 0.782                                & 0.801                                & 0.827                                  \\ \cline{2-7} 
\multirow{-3}{*}{\textbf{JC}}                       & \textbf{Inter} & 0.5                                    & 0.5                                & 0.541                                & 0.502                                & 0.578                                  \\ \hline \hline
                                                    & \textbf{Total} & 0.747                                  & 0.766                              & 0.752                                & 0.746                                & 0.749                                  \\ \cline{2-7} 
                                                    & \textbf{Intra} &  0.751           & 0.790        &  0.785         &  0.772         & 0.796           \\ \cline{2-7} 
\multirow{-3}{*}{\textbf{AA}}                       & \textbf{Inter} & 0.642                                  & 0.568                              & 0.617                                & 0.643                                & 0.641                                  \\ \hline \hline
                                                    & \textbf{Total} & 0.751                                  & 0.770                               & 0.752                                & 0.744                                & 0.751                                  \\ \cline{2-7} 
                                                    & \textbf{Intra} & 0.754                                  & 0.795                              & 0.782                                & 0.786                                & 0.799                                  \\ \cline{2-7} 
\multirow{-3}{*}{\textbf{RA}}                       & \textbf{Inter} & 0.677                                  & 0.568                              & 0.625                                & 0.575                                & 0.643                                  \\ \hline \hline
                                                    & \textbf{Total} & 0.796                                  & 0.793                              & 0.797                                & 0.886                                & 0.838                                  \\ \cline{2-7} 
                                                    & \textbf{Intra} & 0.804                                  & 0.819                              & 0.832                                & 0.906                                & 0.876                                  \\ \cline{2-7} 
\multirow{-3}{*}{\textbf{DeepWalk}}                 & \textbf{Inter} & 0.587                                  & 0.582                              & 0.652                                & 0.807                                & 0.754                                  \\ \hline \hline
%
                                                    & \textbf{Total} & 0.829                                  & 0.820                              & 0.830                                 & 0.900                                & 0.865                                  \\ \cline{2-7} 
                                                    & \textbf{Intra} & 0.834                                  & 0.849                              & 0.864                                 & 0.916                                & 0.897                                  \\ \cline{2-7} 
\multirow{-3}{*}{\textbf{Node2Vec}}                 & \textbf{Inter} & 0.619                                  & 0.582                              & 0.691                                & 0.834                                & 0.793                                  \\ \hline \hline
                                                    & \textbf{Total} & 0.752                                  & 0.806                              & 0.729                                 & 0.865                                & 0.847                                  \\ \cline{2-7} 
                                                    & \textbf{Intra} & 0.765                                  & 0.823                              & 0.733                                 & 0.902                                & 0.905                                  \\ \cline{2-7} 
\multirow{-3}{*}{\textbf{Splitter}}                 & \textbf{Inter} & 0.433                                  & 0.666                              & 0.712                                & 0.718                                & 0.718                                 \\  \hline \hline
                                                    & \textbf{Total} & 0.548                                  & 0.544                              & 0.549                                 & 0.581                                & *                                  \\ \cline{2-7} 
                                                    & \textbf{Intra} & 0.550                                  & 0.546                              & 0.553                                 & 0.581                                & *                                  \\ \cline{2-7} 
\multirow{-3}{*}{\textbf{NECS}}                 & \textbf{Inter} & 0.509                                  & 0.527                              & 0.533                                & 0.580                                & *                                 \\  \hline \hline

\textbf{NodeSim}                                                    & \textbf{Total} & 0.758                                  & 0.836                              & 0.596                                & 0.837                                & 0.690                                  \\ \cline{2-7} 
\textbf{(Average)}                                                    & \textbf{Intra} & 0.758                                  & 0.838                              & 0.591                                & 0.839                                & 0.694                                  \\ \cline{2-7} 
& \textbf{Inter} & 0.756                                   & 0.829                              & 0.617                                 & 0.826                                & 0.681                                  \\ \hline \hline
\textbf{NodeSim}                                                    & \textbf{Total} & 0.824                                  & 0.816                              & 0.784                                & 0.876                                & 0.829                                  \\ \cline{2-7} 
\textbf{(Weighted-L1) }                                                   & \textbf{Intra} & 0.833                                   & 0.845                              & 0.827                                 & 0.911                                & 0.866                                   \\ \cline{2-7} 
& \textbf{Inter} & 0.581                                  & 0.582                              & 0.607                                & 0.740                                & 0.747                                  \\ \hline \hline
\textbf{NodeSim}                                                    & \textbf{Total} & 0.827                                  & 0.833                              & 0.783                                & 0.875                                & 0.831                                  \\ \cline{2-7} 
\textbf{(Weighted-L2)}                                                    & \textbf{Intra} & 0.834                                  & 0.863                              & 0.820                                 & 0.908                                & 0.864                                  \\ \cline{2-7} 
& \textbf{Inter} & 0.651                                  & 0.589                              & 0.628                                & 0.745                                & 0.758                                  \\ \hline \hline
\textbf{NodeSim} & \textbf{Total} & \textbf{0.857}                         & \textbf{0.864}                     & \textbf{0.849}                       & \textbf{0.924}                       & \textbf{0.883}                          \\ \cline{2-7} 
\textbf{(Hadamard)}                                                    & \textbf{Intra} & \textbf{0.862}                         & \textbf{0.874}                     & \textbf{0.884}                       & \textbf{0.937}                       & \textbf{0.907}                         \\ \cline{2-7} 
& \textbf{Inter} & \textbf{0.736}                                  & \textbf{0.706}                             & \textbf{0.749}                                & \textbf{0.872}                       & \textbf{0.828}                         \\ \hline
\end{tabular}
}
\end{table}

\begin{figure}
  \begin{subfigure}{0.5\textwidth}
    \includegraphics[width=\linewidth]{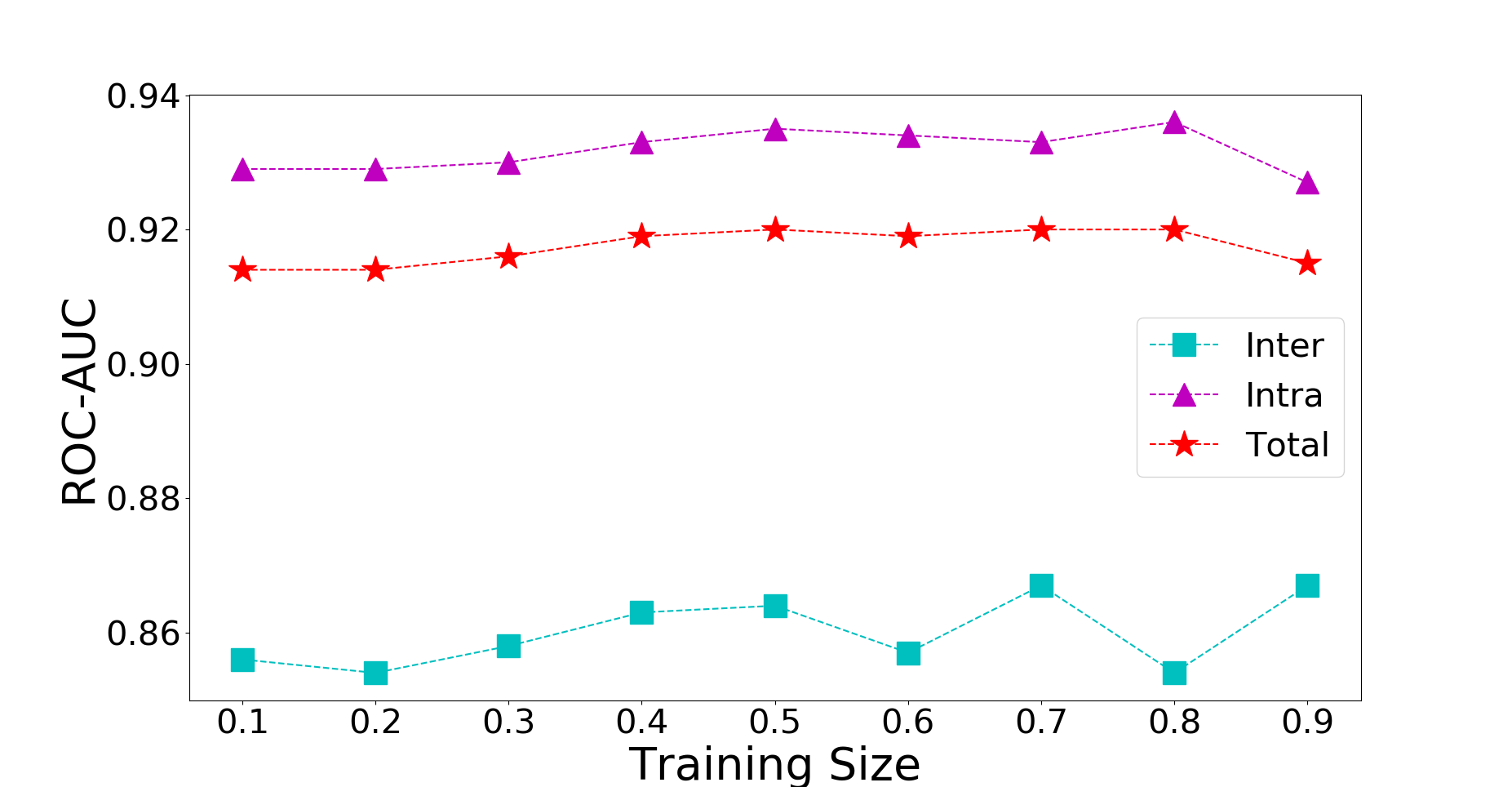}
    \caption{Hep-ph}
    \label{fig:figure1}
  \end{subfigure}%
  \hfill
  \begin{subfigure}{0.5\textwidth}
    \includegraphics[width=\linewidth]{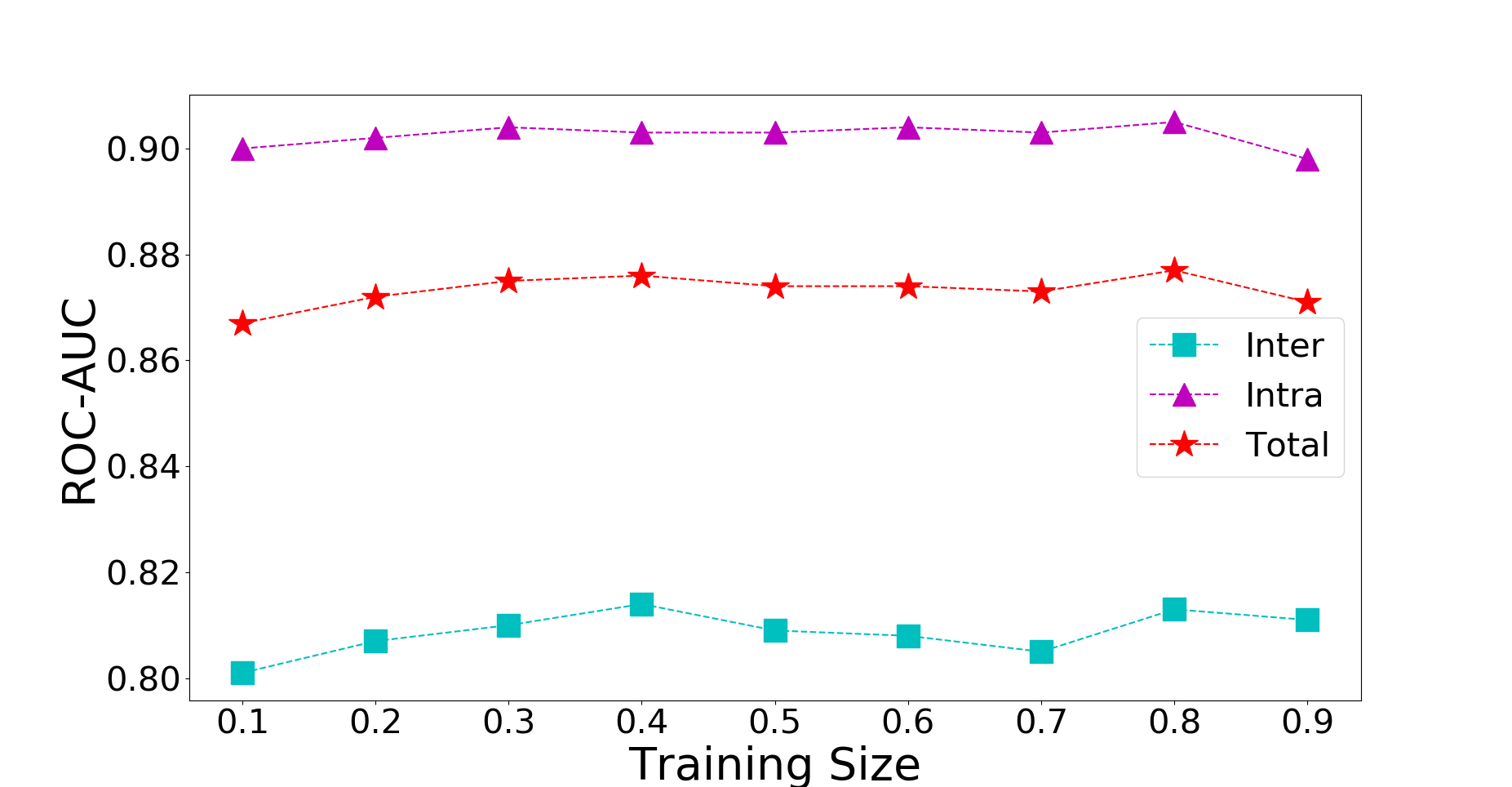}
    \caption{Astro-ph}
    \label{fig:figure2}
  \end{subfigure}%
  \caption{Vary train size.}
  \label{varytrainsize}
\end{figure}

We further study the performance of our method by varying the ratio of train and test set. The results are shown in Figure \ref{varytrainsize} for Hep-ph and Astro-ph networks. Results show that the performance of the proposed method is better compared to baselines, even if the training ratio is $0.1$; however, the best results are achieved when the ratio of training size is at least $0.5$ and $0.3$ for Hep-ph and Astro-ph networks, respectively. 

\subsection{Parameter sensitivity}

The NodeSim embedding method depends on a number of parameters, and we examine the impact of different parameters on the performance of link prediction. In Table \ref{parameters}, we have shown the default values of different parameters and their range that we have considered. The results are shown on the two largest networks, Hep-ph and Astro-ph. 

\begin{table}[b]
\centering
\caption{Default and varied range values of different network embedding parameters.}
\label{parameters}
\resizebox{\textwidth}{!}{
\begin{tabular}{|l|l|l|}
\hline
\textbf{Parameter}       & \textbf{Default} & \textbf{Range}             \\ \hline
$\alpha$        & 1     & 1, 1.5, 2, 2.5, 3 \\ \hline
$\beta$               & 1.5     & 1, 1.5, 2, 2.5, 3 \\ \hline
Dimension ($d$)   & 128     & 4, 8, 16, 32, 64, 128, 256           \\ \hline
Context  ($k$)       & 5       & 5, 7, 9, 11, 13, 15             \\ \hline
Number of Walks ($\gamma$) & 10      &   6, 8, 10, 12, 14, 16, 18, 20                \\ \hline
Walk Length ($l$) & 80      &     40, 50, 60, 70, 80, 90, 100              \\ \hline
\end{tabular}
}
\end{table}

\begin{figure}
  \begin{subfigure}{0.5\textwidth}
    \includegraphics[width=\linewidth]{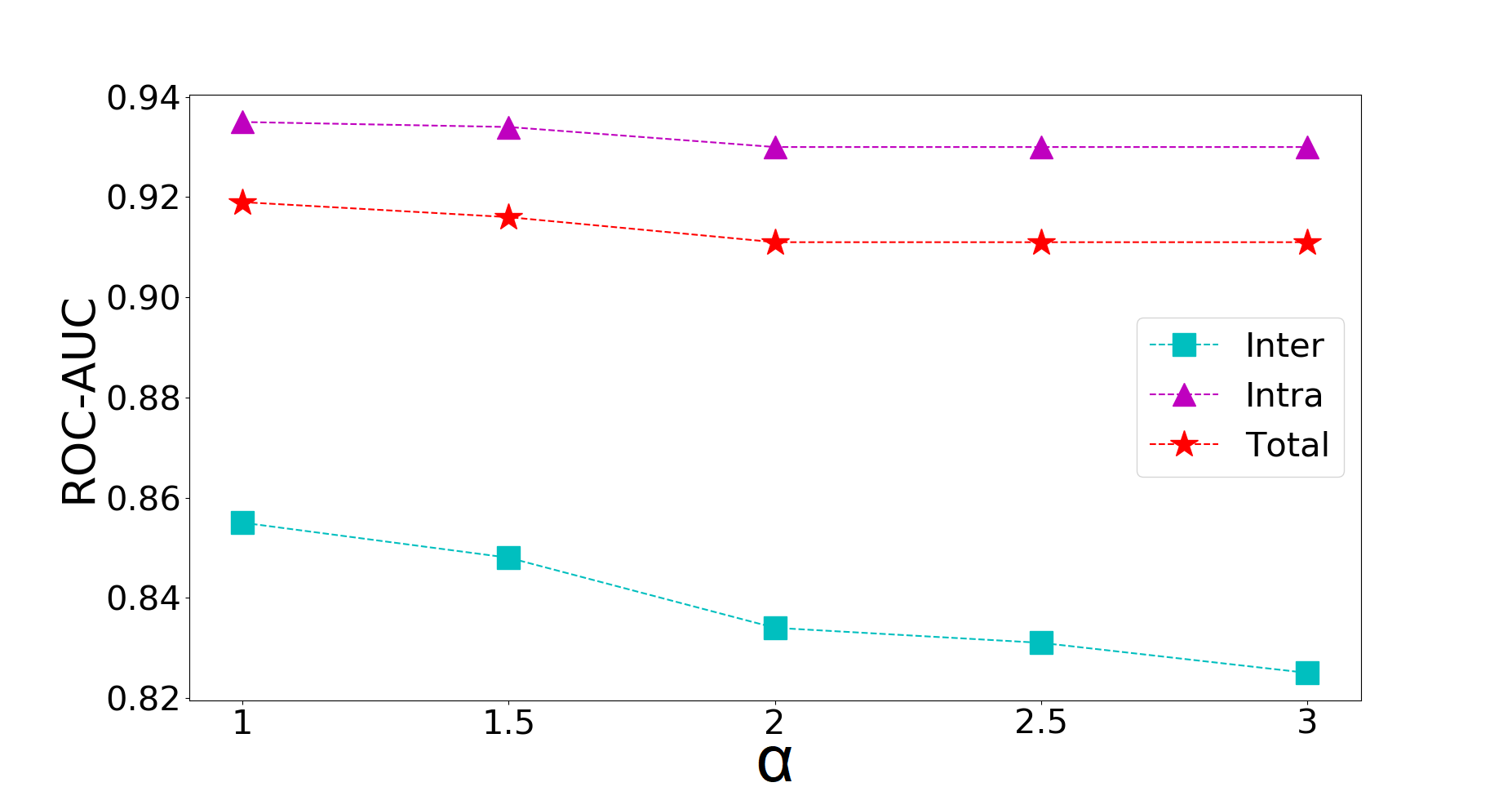}
    \caption{Hep-ph}
    \label{fig:figure1}
  \end{subfigure}%
  \hfill
  \begin{subfigure}{0.5\textwidth}
    \includegraphics[width=\linewidth]{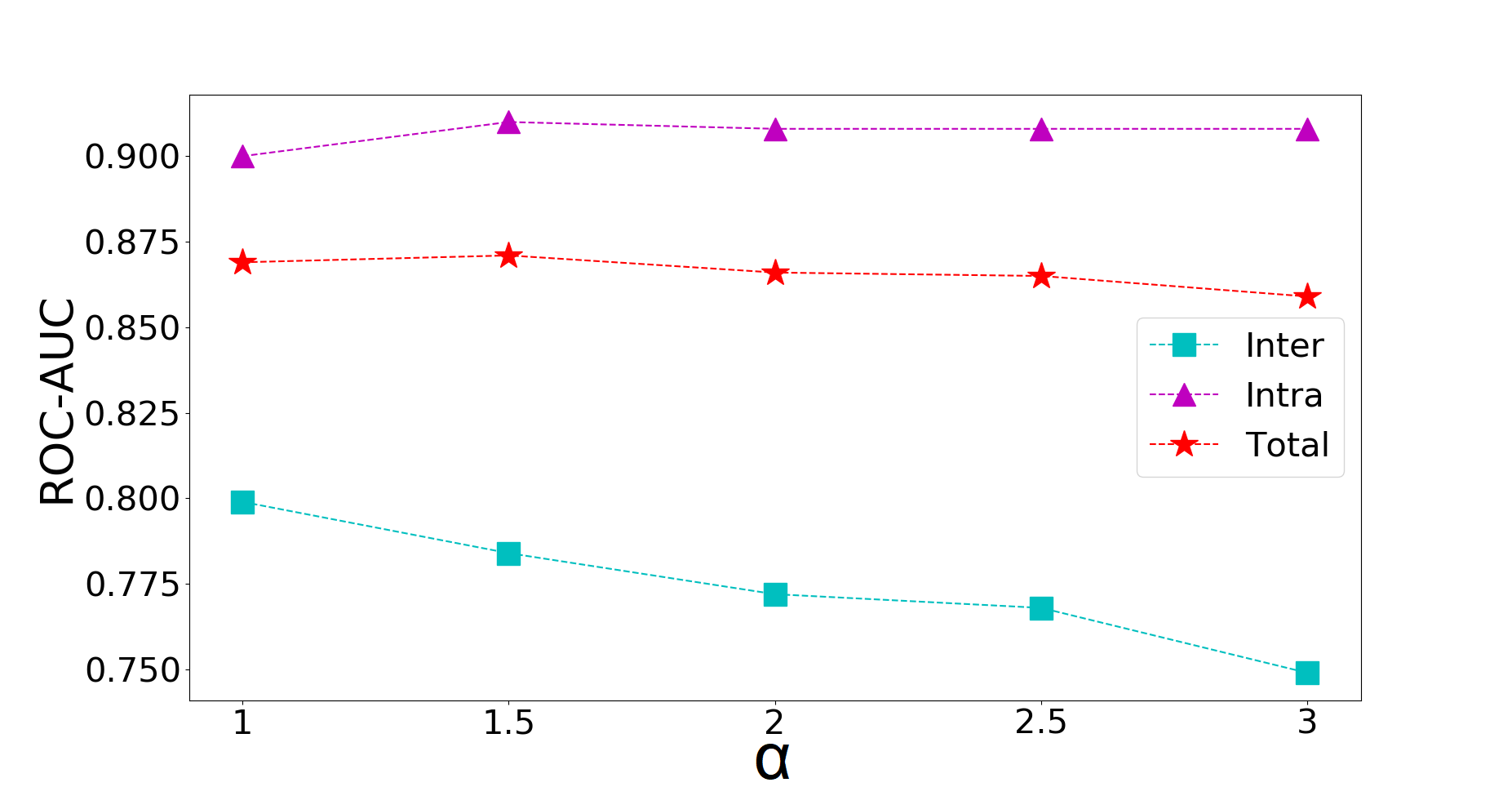}
    \caption{Astro-ph}
    \label{fig:figure2}
  \end{subfigure}%
  \caption{Impact of varying $\alpha$.}
  \label{varya}
\end{figure}

\begin{figure}
  \begin{subfigure}{0.5\textwidth}
    \includegraphics[width=\linewidth]{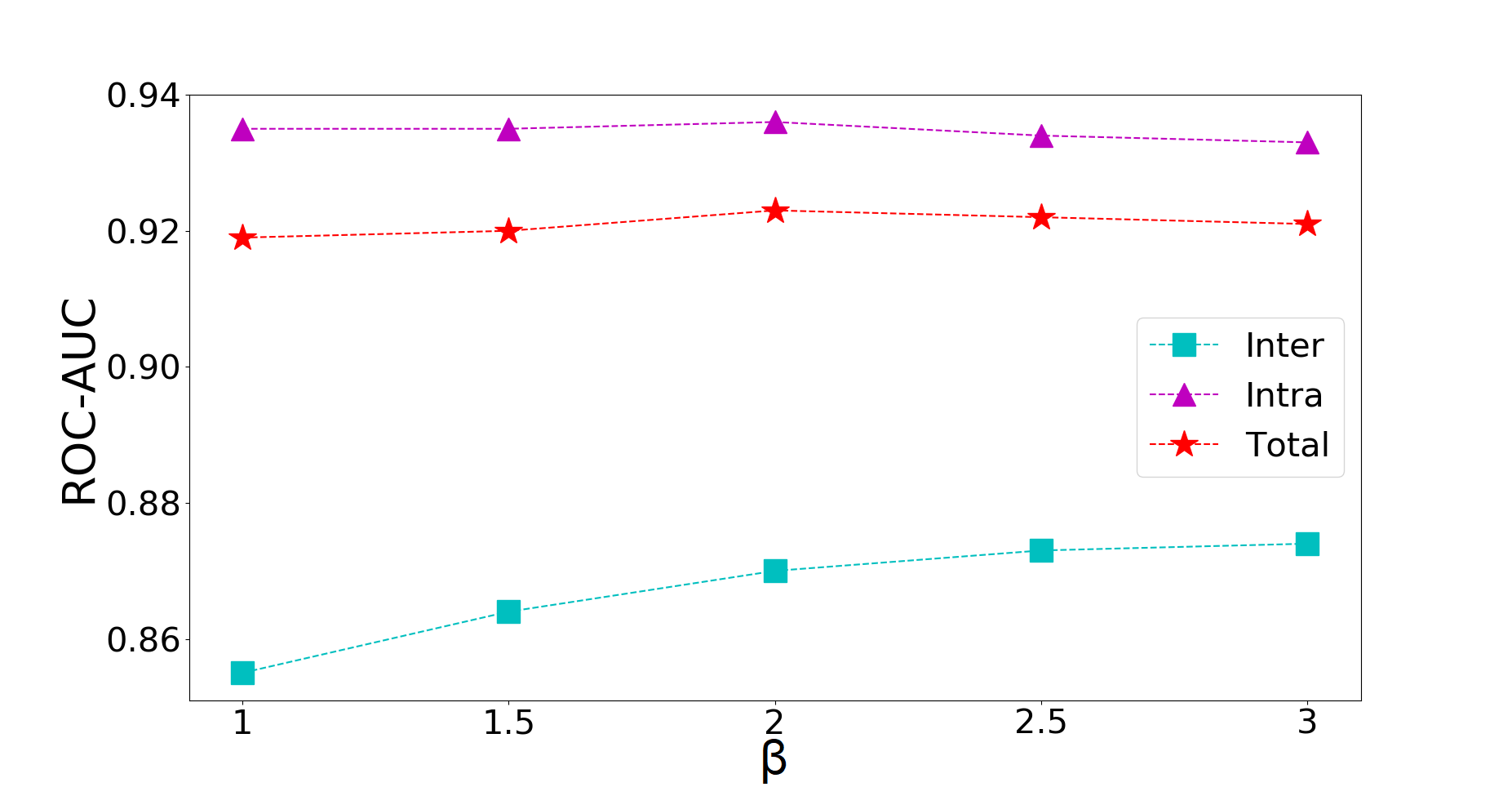}
    \caption{Hep-ph}
    \label{fig:figure1}
  \end{subfigure}%
  \hfill
  \begin{subfigure}{0.5\textwidth}
    \includegraphics[width=\linewidth]{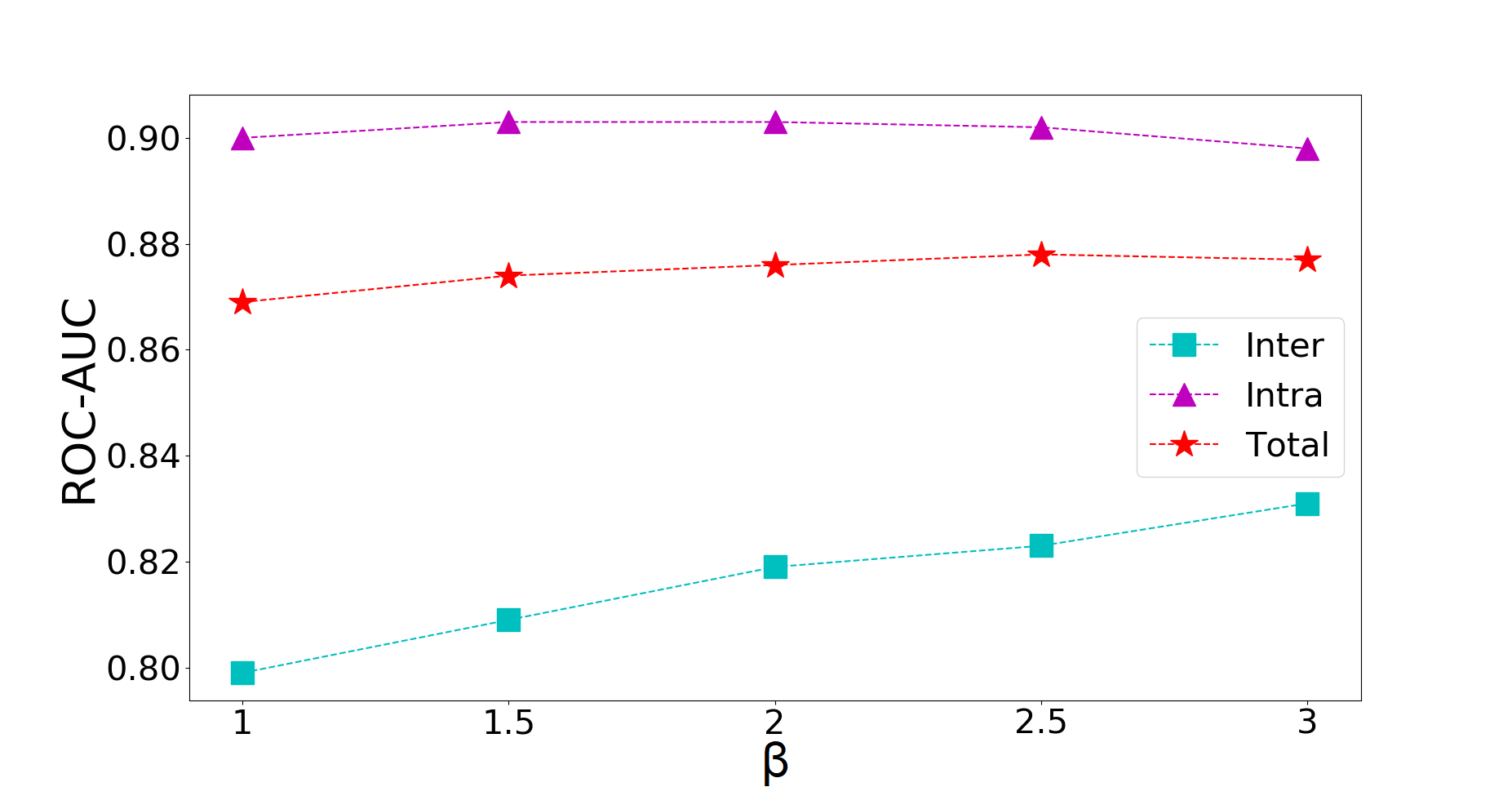}
    \caption{Astro-ph}
    \label{fig:figure2}
  \end{subfigure}%
  \caption{Impact of varying $\beta$.}
  \label{varyb}
\end{figure}

Figure \ref{varya} shows the impact of varying $\alpha$ on inter and intra-community link prediction. The results show that $\alpha \sim 1.5$ achieves the best results. In Figure \ref{varyb}, the results show that $\beta \sim  1.5 - 2$ achieves the best results. The results confirm that the inter-community edges should be weighted higher than the intra-community edges during the sampling to predict inter-community links with high accuracy, as expected.

\begin{figure}
  \begin{subfigure}{0.5\textwidth}
    \includegraphics[width=\linewidth]{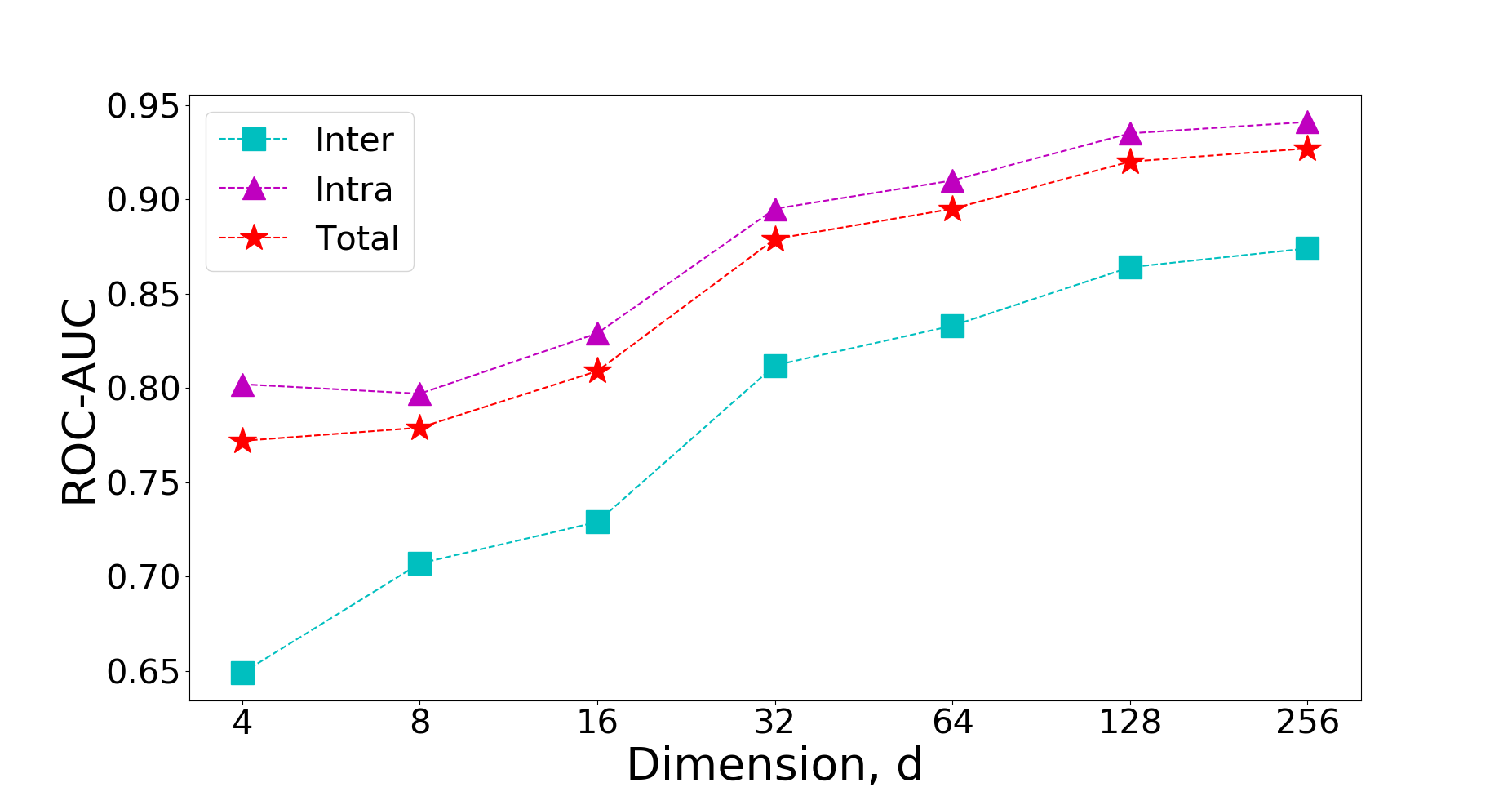}
    \caption{Hep-ph}
    \label{fig:figure1}
  \end{subfigure}%
  \hfill
  \begin{subfigure}{0.5\textwidth}
    \includegraphics[width=\linewidth]{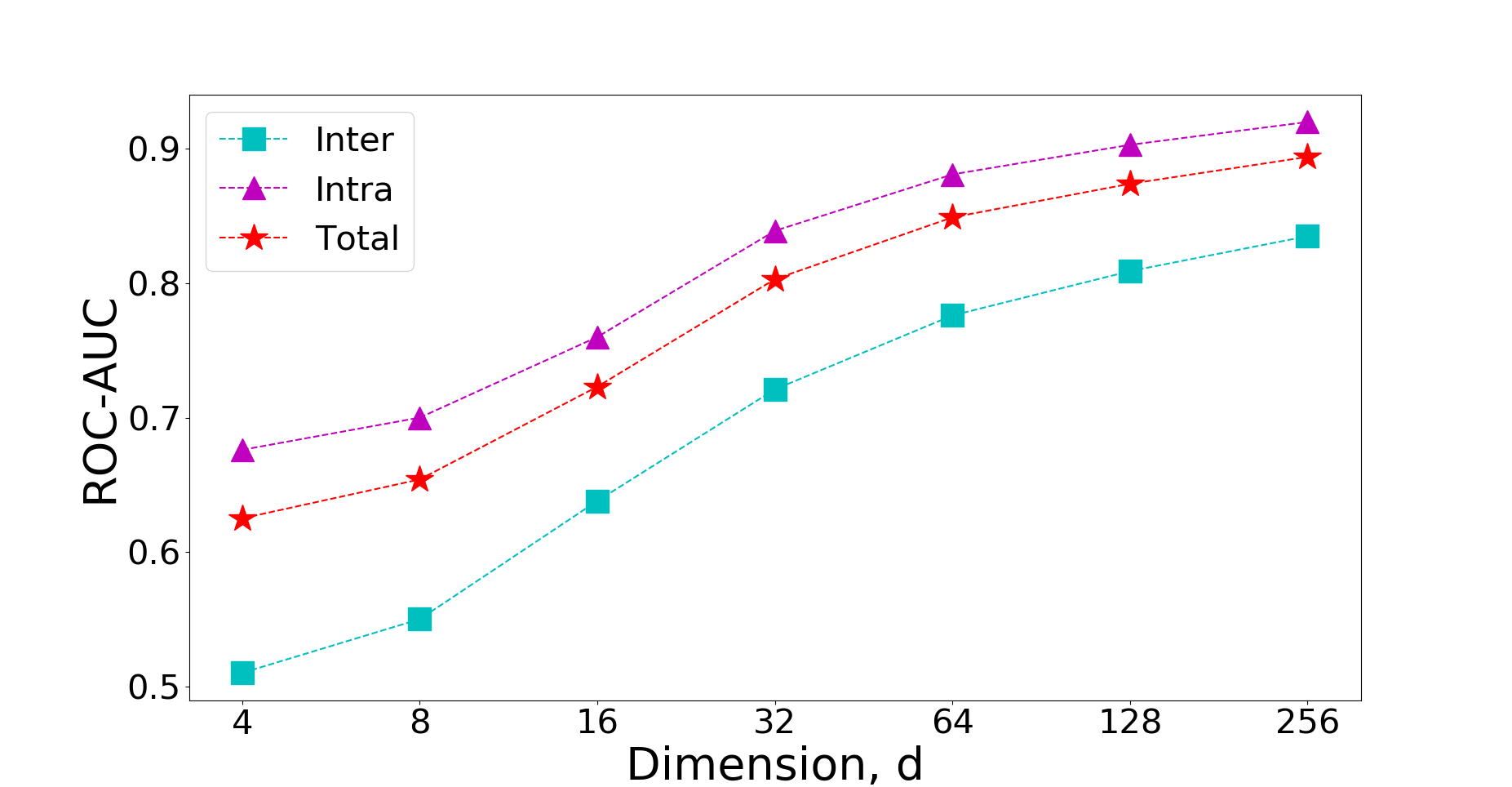}
    \caption{Astro-ph}
    \label{fig:figure2}
  \end{subfigure}%
  \caption{Impact of varying Dimension (d).}
  \label{varydim}
\end{figure}

\begin{figure}
  \begin{subfigure}{0.5\textwidth}
    \includegraphics[width=\linewidth]{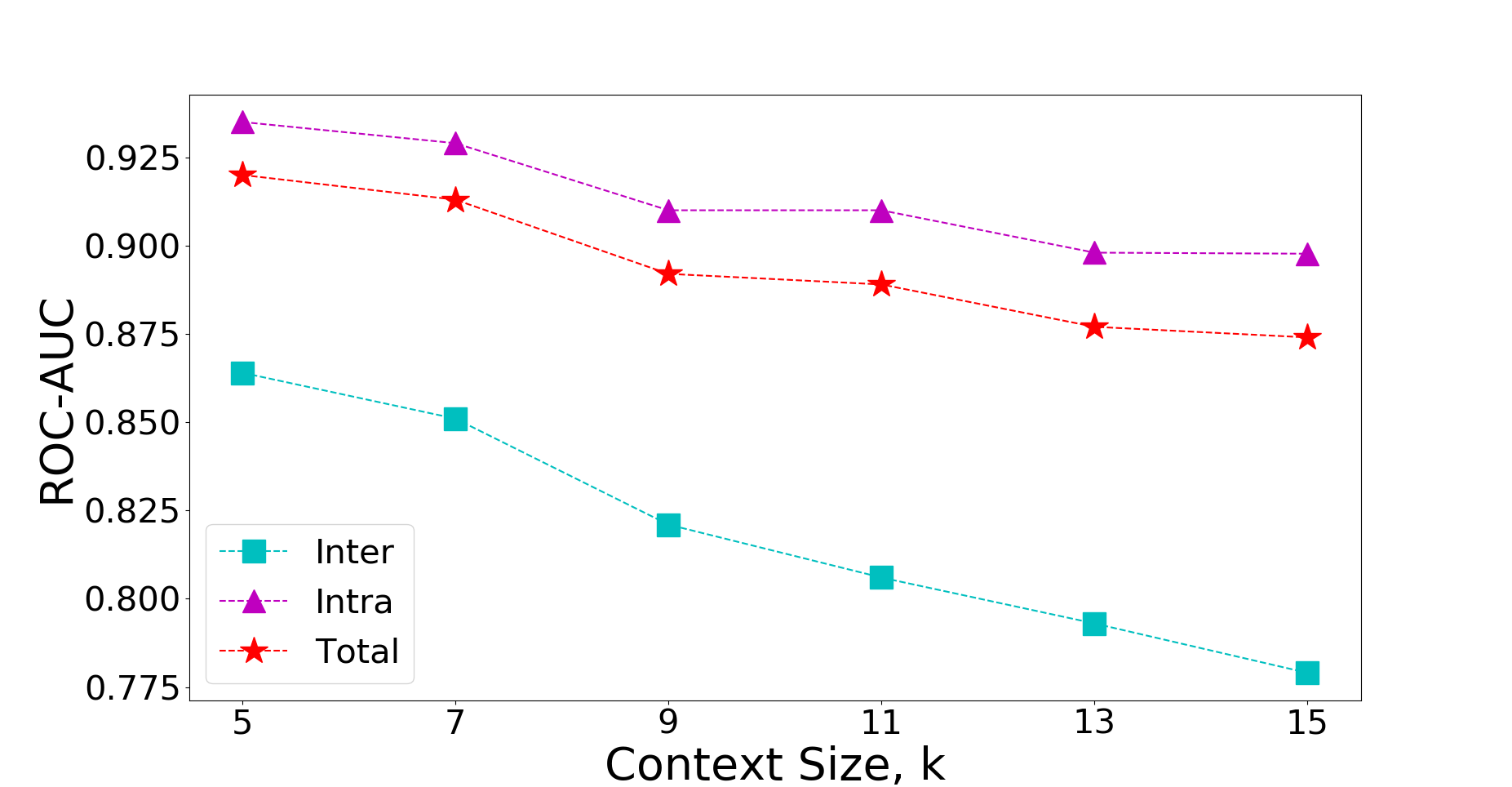}
    \caption{Hep-ph}
    \label{fig:figure1}
  \end{subfigure}%
  \hfill
  \begin{subfigure}{0.5\textwidth}
    \includegraphics[width=\linewidth]{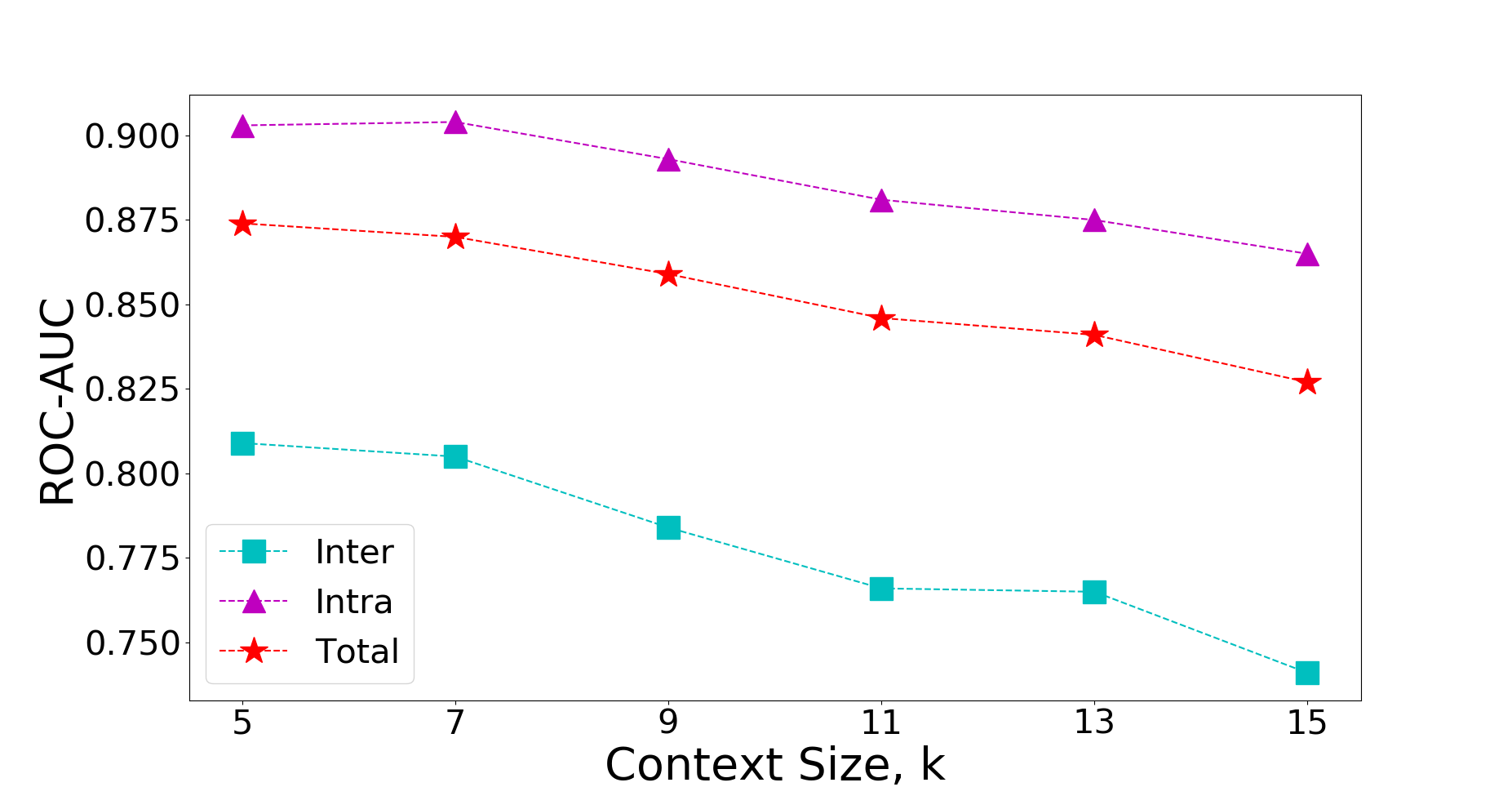}
    \caption{Astro-ph}
    \label{fig:figure2}
  \end{subfigure}%
  \caption{Impact of varying context k.}
  \label{varyws}
\end{figure}

Next, we analyze the impact of embedding parameters on link prediction accuracy. Figure \ref{varydim} represents that the performance of link prediction methods improves with the embedding dimension. In Figure \ref{varyws}, we observe that the performance reduces with the window size as the larger window size considers distant nodes while generating the local context of the nodes, and these nodes might not be similar. In real-world networks, most of the new links are driven by the triad closure phenomenon, and it is less probable that a node will be connected to a distant node. 

\begin{figure}
  \begin{subfigure}{0.5\textwidth}
    \includegraphics[width=\linewidth]{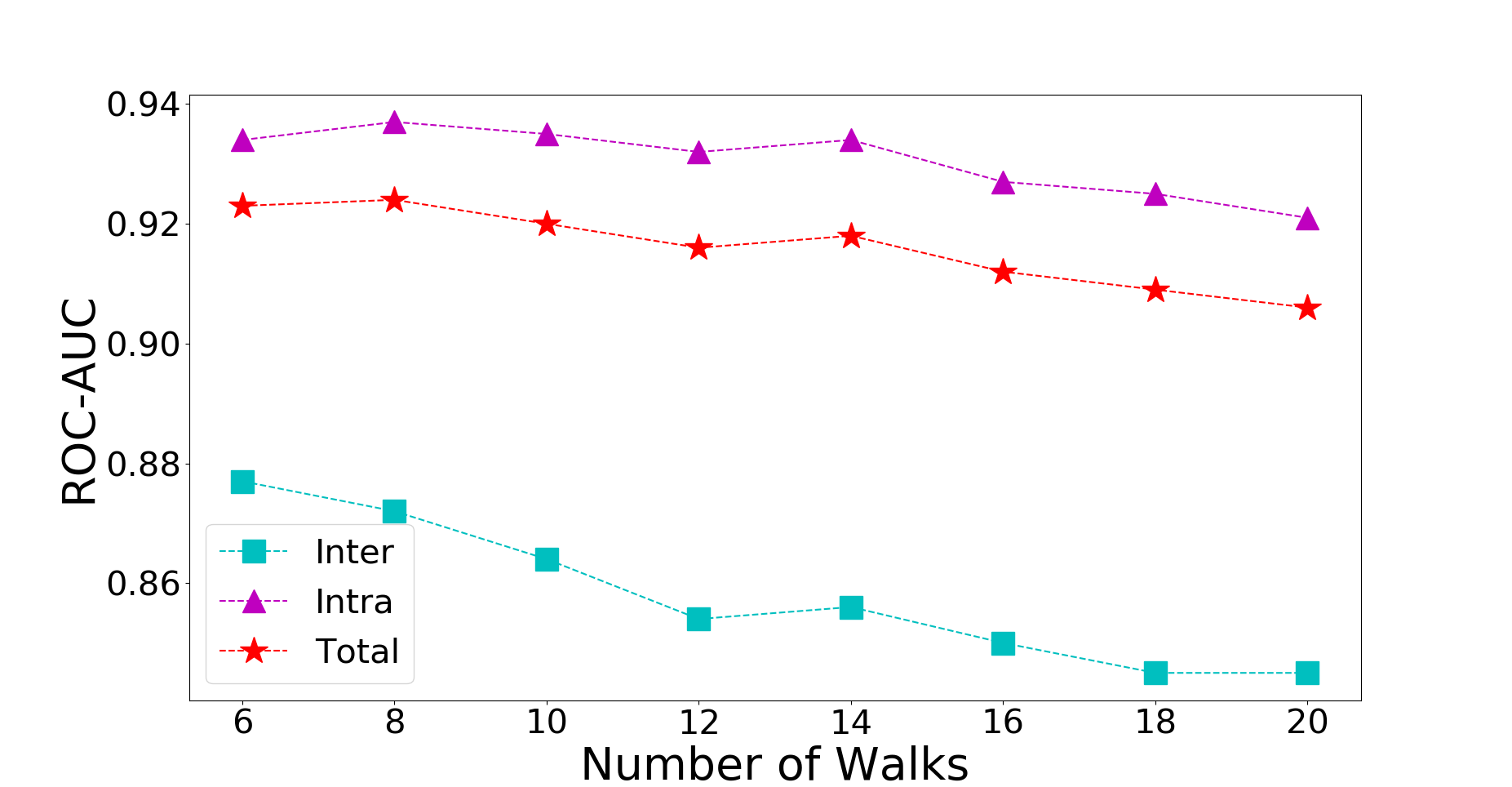}
    \caption{Hep-ph}
    \label{fig:figure1}
  \end{subfigure}%
  \hfill
  \begin{subfigure}{0.5\textwidth}
    \includegraphics[width=\linewidth]{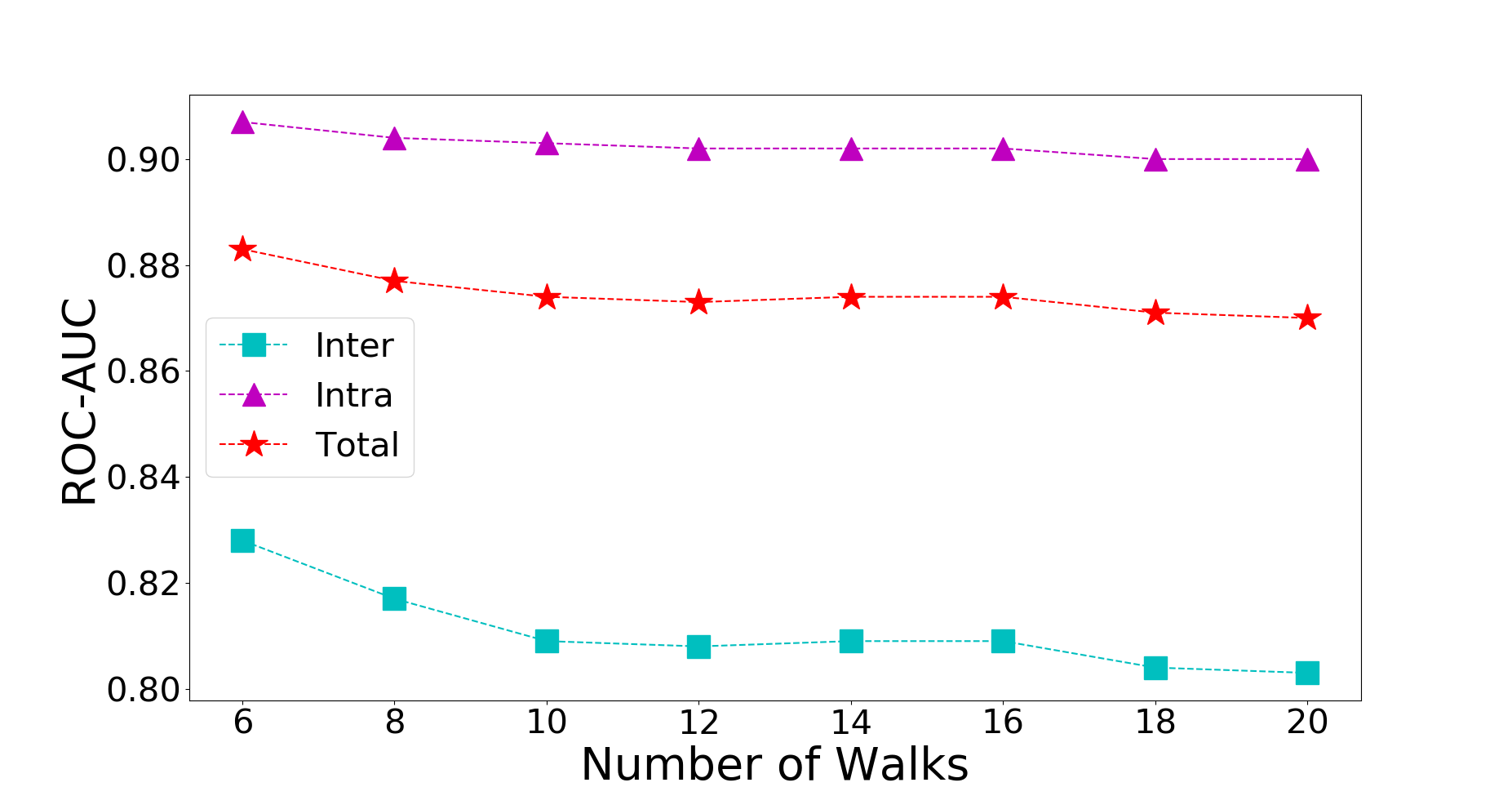}
    \caption{Astro-ph}
    \label{fig:figure2}
  \end{subfigure}%
  \caption{Impact of varying Number of Walks.}
  \label{varynw}
\end{figure}

\begin{figure}
  \begin{subfigure}{0.5\textwidth}
    \includegraphics[width=\linewidth]{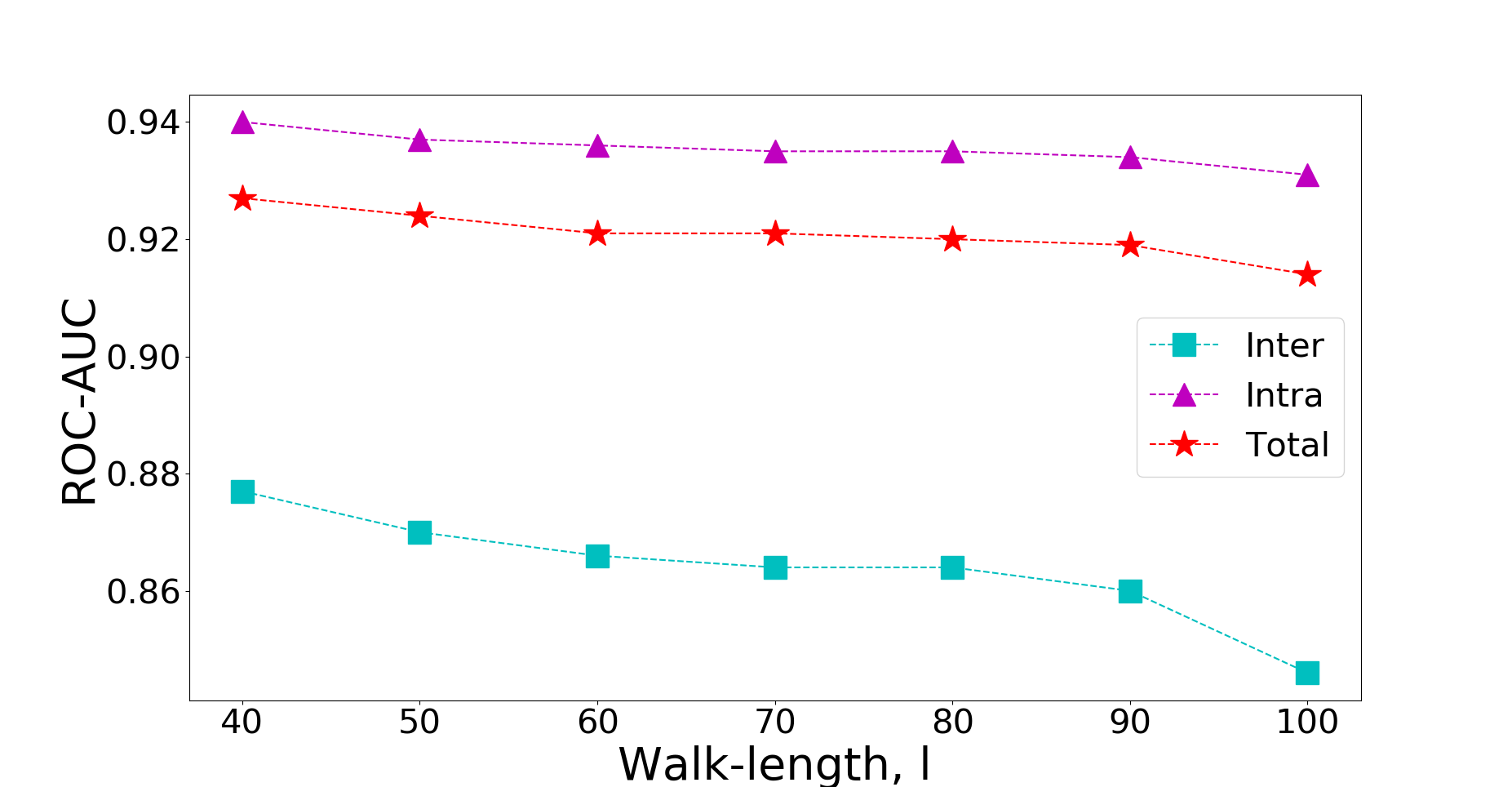}
    \caption{Hep-ph}
    \label{fig:figure1}
  \end{subfigure}%
  \hfill
  \begin{subfigure}{0.5\textwidth}
    \includegraphics[width=\linewidth]{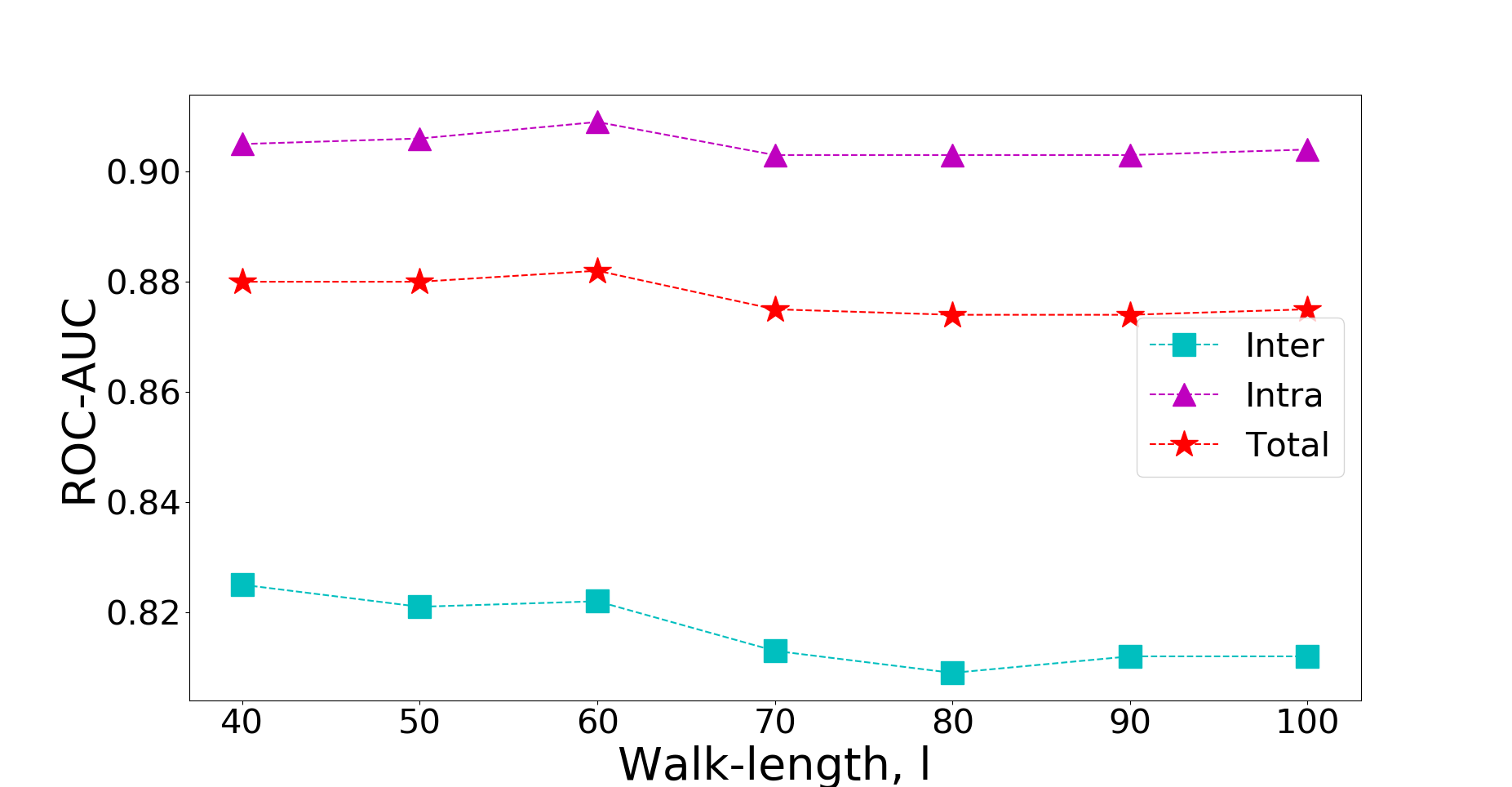}
    \caption{Astro-ph}
    \label{fig:figure2}
  \end{subfigure}%
  \caption{Impact of varying Walk-length.}
  \label{varywl}
\end{figure}

Figures \ref{varynw} and \ref{varywl} show results for varying the number of walks and the walk-length. As observed in Figure \ref{varynw}, the intra-community results are less affected by the number of walks than the inter-community links as the ratio of inter-community context pairs decreases with more number of walks; as we expected. Similarly, the inter-community accuracy also decreases with the walk-length even if the total accuracy is improved, as shown in Figure \ref{varywl} (b).

\subsection{Scalability}

\begin{figure}[t]
\centering
\includegraphics[width=\linewidth]{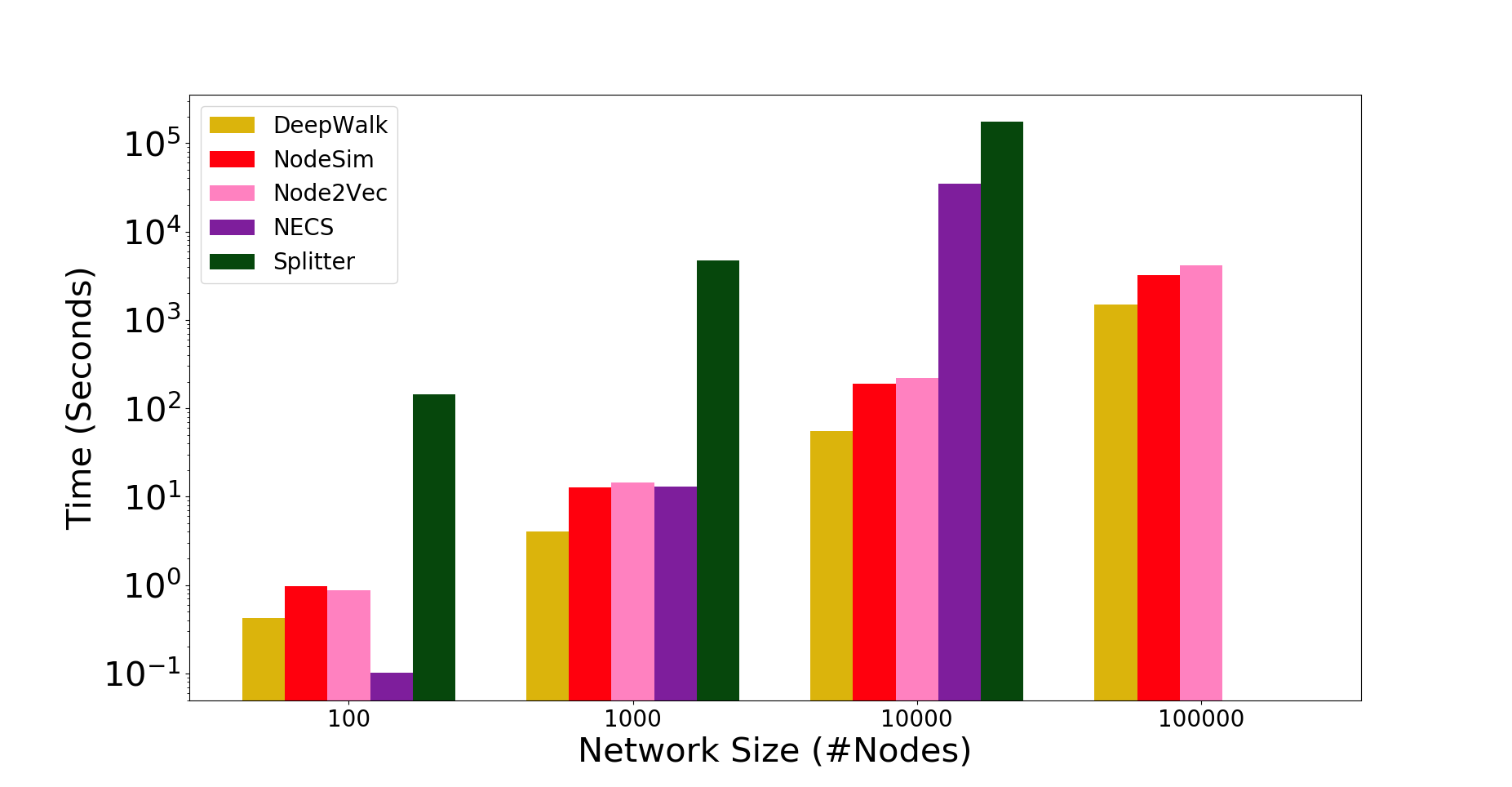}
\caption{Running time for different embedding methods versus network size}
\label{scale}
\end{figure}

We compare the running time of different network embedding based methods on synthetic networks generated using SCCP (Scale-free networks with Community and Core-Periphery) model \citep{saxena2015understanding,gupta2016modeling}. The network generator first creates a seed graph, i.e., a complete graph of $m$ nodes for each community, where $m$ is the average degree of nodes. Next, in each iteration, a new node is added to each community, and the added node builds $m$ connections using preferential attachment law \citep{barabasi1999emergence} while ensuring the intra and inter-community edge ratio. The running time is compared on synthetic networks so that the ratio of intra and inter-community edges are maintained as we increase the network size. In our experiments, the ratio is (intra : inter= $.75:.25$), and the average degree of the network is $8$. The total number of communities is $10$ in the network having $100$ and $1000$ nodes and $100$ in the network having $10000$ and $100000$ nodes. All communities in a network are of the same size. 

Figure \ref{scale} show the running time of different methods. All experiments are performed on the server having 384GB RAM and 2x Intel Xeon 4110 @ 2.1Ghz CPU. For $100000$ nodes network, the Splitter code was not finished in 48 hours, and the NECS code was killed due to the memory error on the server. 
The results show that the proposed method executes faster than all the baselines except deepWalk as the network size grows. The deepwalk method is the fastest as it creates node context using a simple random walk and does not consider the structural properties of the network.

\subsection{Robustness for Identified Communities}

There have been proposed several community detection methods in the literature that consider different network properties while identifying the communities. Therefore, the communities identified by different methods might vary. For some methods, such as Louvain or greedy method, if the same method is applied many times, the returned community structure might differ each time.

We, therefore, study the efficiency of the NodeSim embedding method corresponding to different community detection methods. We apply five different community detection methods (including Louvain), which are mentioned below.

\begin{enumerate}
    \item Asynchronous Label Propagation \citep{raghavan2007near}: In this method, each node is initialized with a unique community label. In every iteration, each node will update its community label based on its neighbors' community label at that time. Thus, the nodes belonging to a strongly connected group will be assigned the same community label with their consensus through this iterative process. 
     
    \item Semi-synchronous Label Propagation \citep{cordasco2010community}: This method is similar to the asynchronous Label propagation method, and it combines the advantages of both synchronous and asynchronous method. In this method, each node is assigned with a community label initially, and at each iteration, a node updates its community label based on the most used label by its neighbors. However, the ties are broken randomly, and the method is stopped when no node changes its label. 
    
    \item Fluid Communities Algorithm \citep{pares2017fluid}: Fluid communities are based on the idea of fluids interacting with each other, such as expanding or pushing each other in an environment. In this method, first, each of the initial $c$ communities is initialized by a random node in the network. Then, in each iteration, each node's community label is updated based on its community and the community of its neighbors. Once no node changes its community in an iteration, the method is stopped. 
    In our implementation, we set the number of communities approximately close to the number of communities identified by the Louvain method.

    \item Greedy Modularity Maximization \citep{clauset2004finding}: This method is a well-known method to identify communities by maximizing the modularity in the network. In this method, each node is assigned with a community label, and in each step, two communities are merged that most increases the modularity. The method is stopped when the modularity can not be further increased by merging two communities. 
\end{enumerate}

\begin{figure}[t]
\centering
\includegraphics[width=\linewidth]{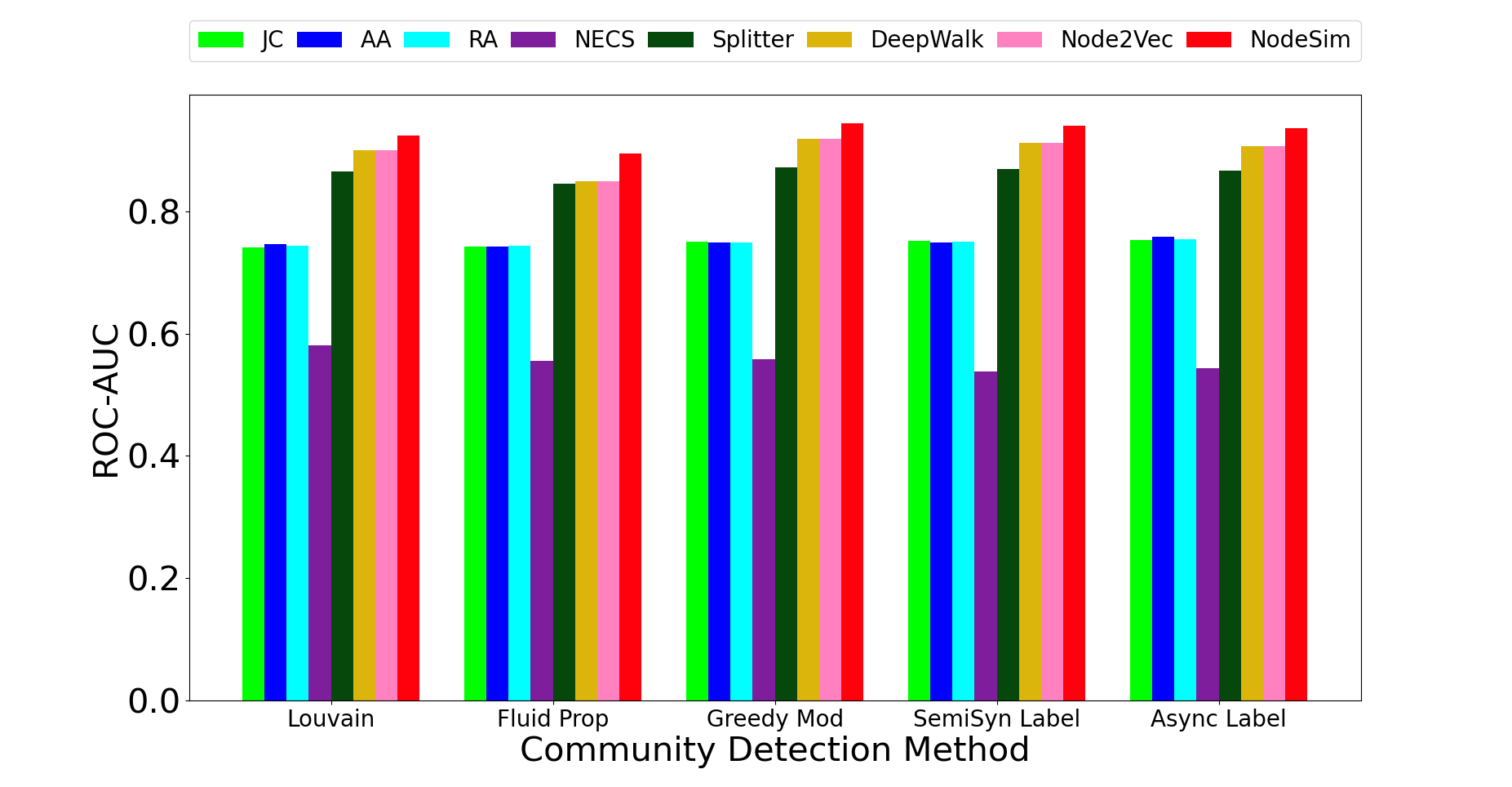}
\caption{ROC-AUC for link prediction corresponding to different community detection methods for Hep-ph network.}
\label{commimpact}
\end{figure}

After identifying the communities using different methods, the training and testing data is created, as discussed in Section \ref{datasets}. Next, we generate network embedding by applying different embedding methods and apply the link prediction method. Each method is executed five times, and the average ROC-AUC value for the Hep-ph network is shown in Figure \ref{commimpact}.

The results show that the performance of different methods is relatively maintained irrespective of the community detection method. The NodeSim method outperforms in all the cases as the method considers both the similarity of nodes and their communities while generating the network embedding.

\subsection{Case Study}

\begin{figure}[t]
  \begin{subfigure}{0.5\textwidth}
    \includegraphics[width=\linewidth]{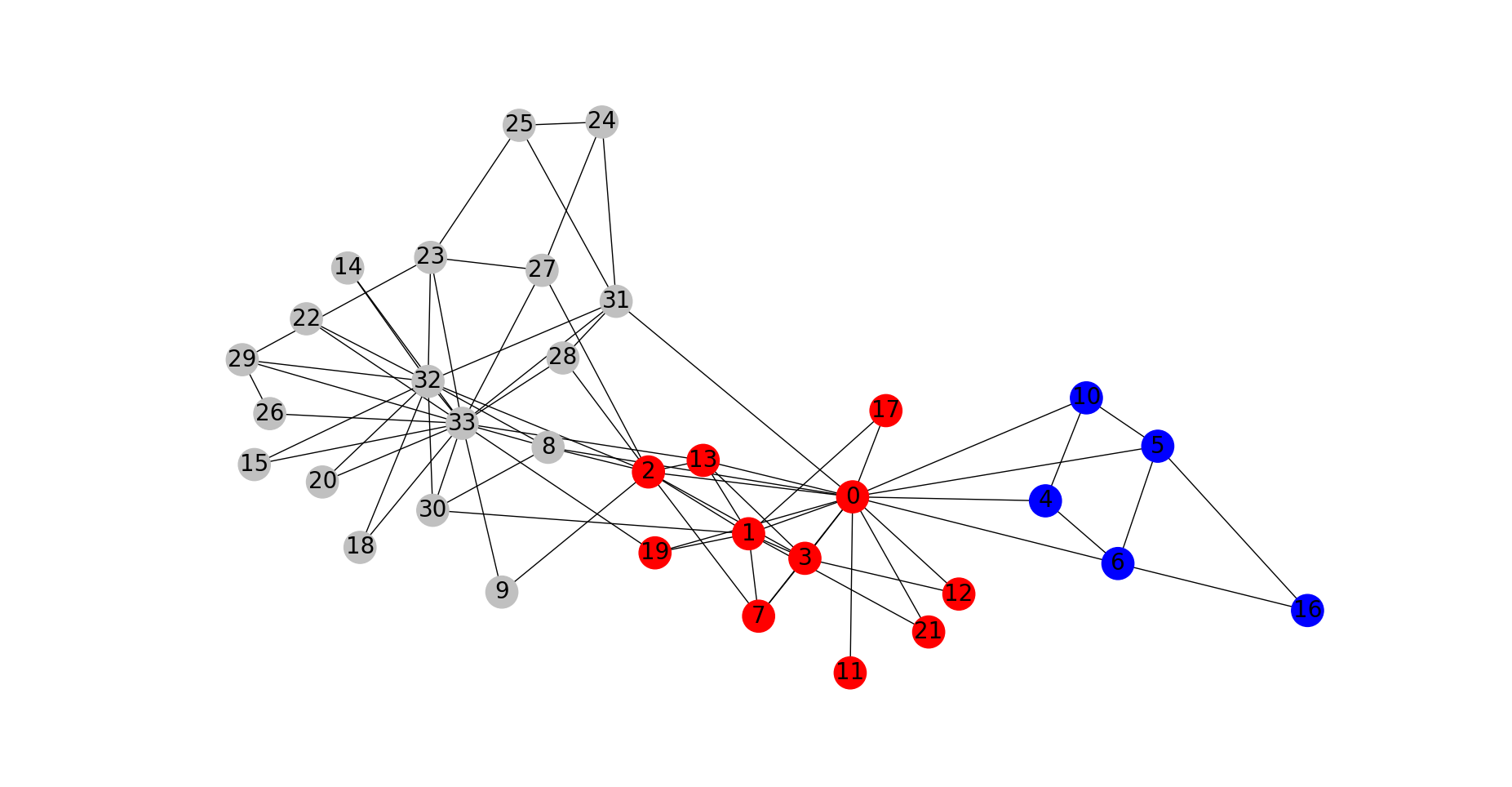}
    \caption{}
    \label{fig:figure1}
  \end{subfigure}%
  \hfill
  \begin{subfigure}{0.5\textwidth}
    \includegraphics[width=\linewidth]{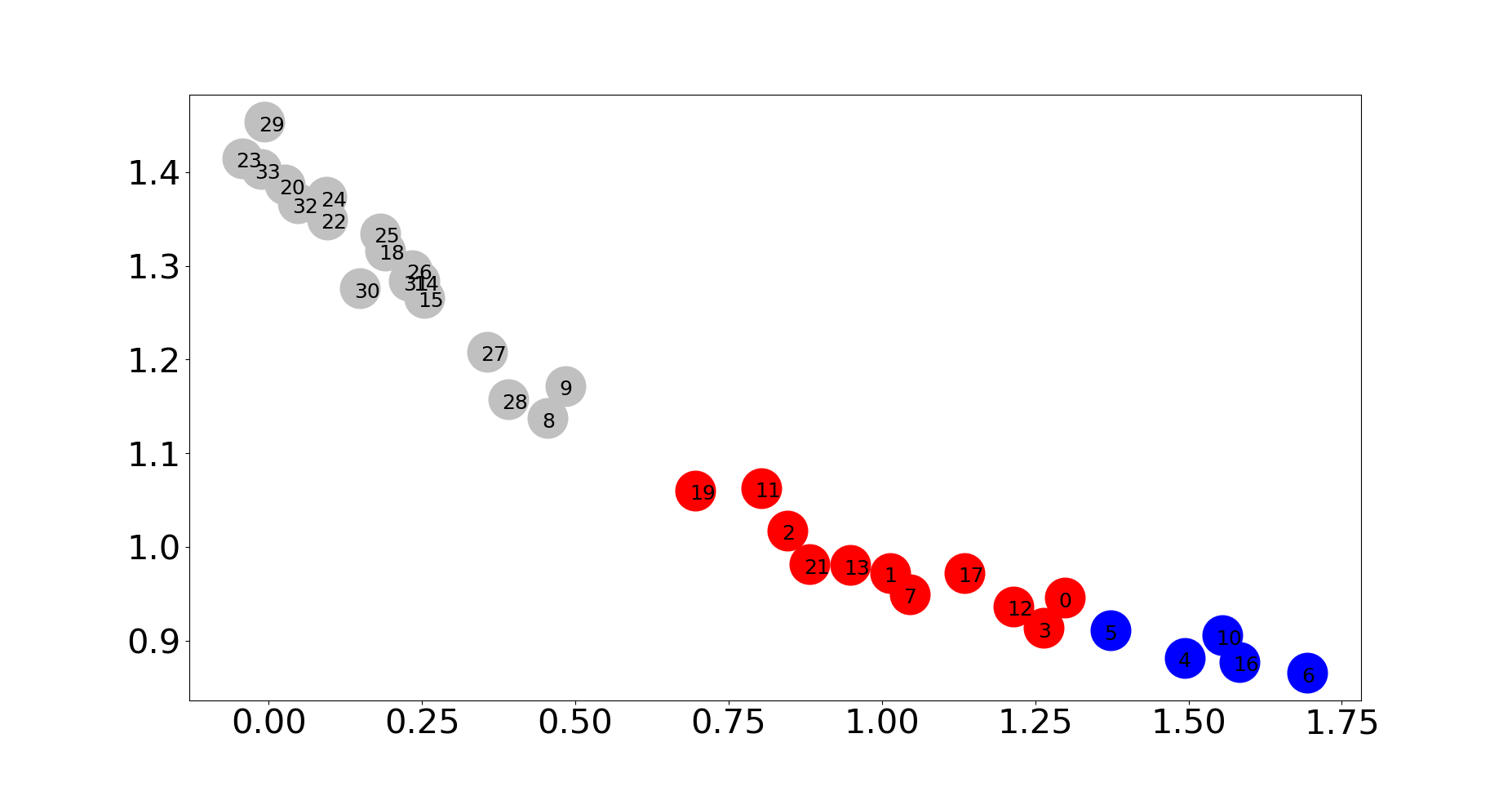}
    \caption{}
    \label{fig:figure2}
  \end{subfigure}%
  \caption{(a) Zachary Karate Network with three communities and (b) 2-dimension embedding of Zachary Karate Network using NodeSim Method.}
  \label{zachary}
\end{figure}

For visualization, we show the NodeSim embedding of the Zachary Karate Network \citep{zachary1977information} in 2-dimension space. The network and its embedding are shown in Figure \ref{zachary}, where the nodes having the same color belong to one community. The embedding shows that the nodes belonging to different communities are well separated; however, more similar nodes are embedded closer. For example, node 12 is more probable to form inter-community links with node 5 and node 4, so, as observed, they are embedded closer but still well separated. 

\section{Conclusion}\label{conclusion}

In this work, we have proposed the NodeSim network embedding method, which considers both the nodes' similarity and their community membership while learning the feature representation of the nodes. The NodeSim embedding method efficiently learns the embedding of diverse nodes that is further verified using the link prediction. We proposed a link prediction method that trains a logistic regression model using nodes' features and their community information. The results showed that the proposed link-prediction method outperforms baseline methods for both intra-community as well as inter-community link prediction. We further studied the impact of different parameters and showed that a higher value of $\beta$ provides higher inter-community link prediction accuracy as the NodeSim method embeds the more similar diverse nodes closer than the others. The proposed method can be directly applied to weighted networks. We will further extend the proposed method to generate embedding of dynamic networks to predict inter and intra-community links with high accuracy. 


%
%

\bibliography{mybib}

\begin{thebibliography}{10}

\bibitem{clauset2004finding}
Aaron Clauset, Mark~EJ Newman, and Cristopher Moore.
\newblock Finding community structure in very large networks.
\newblock {\em Physical review E}, 70(6):066111, 2004.

\bibitem{mcpherson2001birds}
Miller McPherson, Lynn Smith-Lovin, and James~M Cook.
\newblock Birds of a feather: Homophily in social networks.
\newblock {\em Annual review of sociology}, 27(1):415--444, 2001.

\bibitem{granovetter1983strength}
Mark Granovetter.
\newblock The strength of weak ties: A network theory revisited.
\newblock {\em Sociological theory}, pages 201--233, 1983.

\bibitem{saxena2016evolving}
Akrati Saxena and SRS Iyengar.
\newblock Evolving models for meso-scale structures.
\newblock In {\em 2016 8th International Conference on Communication Systems
  and Networks (COMSNETS)}, pages 1--8. IEEE, 2016.

\bibitem{benevenuto2009characterizing}
Fabr{\'\i}cio Benevenuto, Tiago Rodrigues, Meeyoung Cha, and Virg{\'\i}lio
  Almeida.
\newblock Characterizing user behavior in online social networks.
\newblock In {\em Proceedings of the 9th ACM SIGCOMM Conference on Internet
  Measurement}, pages 49--62, 2009.

\bibitem{wilson2009user}
Christo Wilson, Bryce Boe, Alessandra Sala, Krishna~PN Puttaswamy, and Ben~Y
  Zhao.
\newblock User interactions in social networks and their implications.
\newblock In {\em Proceedings of the 4th ACM European conference on Computer
  systems}, pages 205--218, 2009.

\bibitem{saxena2020mitigating}
Akrati Saxena, Wynne Hsu, Mong~Li Lee, Hai Leong~Chieu, Lynette Ng, and Loo~Nin
  Teow.
\newblock Mitigating misinformation in online social network with top-k
  debunkers and evolving user opinions.
\newblock In {\em Companion Proceedings of the Web Conference 2020}, pages
  363--370, 2020.

\bibitem{masrour2020bursting}
Farzan Masrour, Tyler Wilson, Heng Yan, Pang-Ning Tan, and Abdol Esfahanian.
\newblock Bursting the filter bubble: Fairness-aware network link prediction.
\newblock In {\em Proceedings of the AAAI Conference on Artificial
  Intelligence}, volume~34, pages 841--848, 2020.

\bibitem{aslay2018maximizing}
Cigdem Aslay, Antonis Matakos, Esther Galbrun, and Aristides Gionis.
\newblock Maximizing the diversity of exposure in a social network.
\newblock In {\em 2018 IEEE International Conference on Data Mining (ICDM)},
  pages 863--868. IEEE, 2018.

\bibitem{zhou2009predicting}
Tao Zhou, Linyuan L{\"u}, and Yi-Cheng Zhang.
\newblock Predicting missing links via local information.
\newblock {\em The European Physical Journal B}, 71(4):623--630, 2009.

\bibitem{liben2007link}
David Liben-Nowell and Jon Kleinberg.
\newblock The link-prediction problem for social networks.
\newblock {\em Journal of the American society for information science and
  technology}, 58(7):1019--1031, 2007.

\bibitem{adamic2003friends}
Lada~A Adamic and Eytan Adar.
\newblock Friends and neighbors on the web.
\newblock {\em Social networks}, 25(3):211--230, 2003.

\bibitem{barabasi1999emergence}
Albert-L{\'a}szl{\'o} Barab{\'a}si and R{\'e}ka Albert.
\newblock Emergence of scaling in random networks.
\newblock {\em science}, 286(5439):509--512, 1999.

\bibitem{valverde2013exploiting}
Jorge Valverde-Rebaza and Alneu de~Andrade~Lopes.
\newblock Exploiting behaviors of communities of twitter users for link
  prediction.
\newblock {\em Social Network Analysis and Mining}, 3(4):1063--1074, 2013.

\bibitem{jeon2017community}
Hyoungjun Jeon and Taewhan Kim.
\newblock Community-adaptive link prediction.
\newblock In {\em Proceedings of the 2017 International Conference on Data
  Mining, Communications and Information Technology}, pages 1--5, 2017.

\bibitem{grover2016node2vec}
Aditya Grover and Jure Leskovec.
\newblock node2vec: Scalable feature learning for networks.
\newblock In {\em Proceedings of the 22nd ACM SIGKDD international conference
  on Knowledge discovery and data mining}, pages 855--864, 2016.

\bibitem{valverde2012structural}
Jorge Valverde-Rebaza and Alneu de~Andrade~Lopes.
\newblock Structural link prediction using community information on twitter.
\newblock In {\em 2012 Fourth International Conference on Computational Aspects
  of Social Networks (CASoN)}, pages 132--137. IEEE, 2012.

\bibitem{clauset2008hierarchical}
Aaron Clauset, Cristopher Moore, and Mark~EJ Newman.
\newblock Hierarchical structure and the prediction of missing links in
  networks.
\newblock {\em Nature}, 453(7191):98--101, 2008.

\bibitem{wang2007local}
Chao Wang, Venu Satuluri, and Srinivasan Parthasarathy.
\newblock Local probabilistic models for link prediction.
\newblock In {\em Seventh IEEE international conference on data mining (ICDM
  2007)}, pages 322--331. IEEE, 2007.

\bibitem{scripps2008matrix}
Jerry Scripps, Pang-Ning Tan, Feilong Chen, and Abdol-Hossein Esfahanian.
\newblock A matrix alignment approach for link prediction.
\newblock In {\em 2008 19th International Conference on Pattern Recognition},
  pages 1--4. IEEE, 2008.

\bibitem{menon2011link}
Aditya~Krishna Menon and Charles Elkan.
\newblock Link prediction via matrix factorization.
\newblock In {\em Joint european conference on machine learning and knowledge
  discovery in databases}, pages 437--452. Springer, 2011.

\bibitem{al2006link}
Mohammad Al~Hasan, Vineet Chaoji, Saeed Salem, and Mohammed Zaki.
\newblock Link prediction using supervised learning.
\newblock In {\em SDM06: workshop on link analysis, counter-terrorism and
  security}, volume~30, pages 798--805, 2006.

\bibitem{lu2010supervised}
Zhengdong Lu, Berkant Savas, Wei Tang, and Inderjit~S Dhillon.
\newblock Supervised link prediction using multiple sources.
\newblock In {\em 2010 IEEE international conference on data mining}, pages
  923--928. IEEE, 2010.

\bibitem{benchettara2010supervised}
Nesserine Benchettara, Rushed Kanawati, and C{\'e}line Rouveirol.
\newblock A supervised machine learning link prediction approach for academic
  collaboration recommendation.
\newblock In {\em Proceedings of the fourth ACM conference on Recommender
  systems}, pages 253--256, 2010.

\bibitem{kashima2009link}
Hisashi Kashima, Tsuyoshi Kato, Yoshihiro Yamanishi, Masashi Sugiyama, and Koji
  Tsuda.
\newblock Link propagation: A fast semi-supervised learning algorithm for link
  prediction.
\newblock In {\em Proceedings of the 2009 SIAM international conference on data
  mining}, pages 1100--1111. SIAM, 2009.

\bibitem{hu2017lpi}
Huan Hu, Chunyu Zhu, Haixin Ai, Li~Zhang, Jian Zhao, Qi~Zhao, and Hongsheng
  Liu.
\newblock Lpi-etslp: lncrna--protein interaction prediction using eigenvalue
  transformation-based semi-supervised link prediction.
\newblock {\em Molecular BioSystems}, 13(9):1781--1787, 2017.

\bibitem{tang2015line}
Jian Tang, Meng Qu, Mingzhe Wang, Ming Zhang, Jun Yan, and Qiaozhu Mei.
\newblock Line: Large-scale information network embedding.
\newblock In {\em Proceedings of the 24th international conference on world
  wide web}, pages 1067--1077, 2015.

\bibitem{wang2016structural}
Daixin Wang, Peng Cui, and Wenwu Zhu.
\newblock Structural deep network embedding.
\newblock In {\em Proceedings of the 22nd ACM SIGKDD international conference
  on Knowledge discovery and data mining}, pages 1225--1234, 2016.

\bibitem{perozzi2014deepwalk}
Bryan Perozzi, Rami Al-Rfou, and Steven Skiena.
\newblock Deepwalk: Online learning of social representations.
\newblock In {\em Proceedings of the 20th ACM SIGKDD international conference
  on Knowledge discovery and data mining}, pages 701--710, 2014.

\bibitem{cao2015grarep}
Shaosheng Cao, Wei Lu, and Qiongkai Xu.
\newblock Grarep: Learning graph representations with global structural
  information.
\newblock In {\em Proceedings of the 24th ACM international on conference on
  information and knowledge management}, pages 891--900, 2015.

\bibitem{du2018galaxy}
Lun Du, Zhicong Lu, Yun Wang, Guojie Song, Yiming Wang, and Wei Chen.
\newblock Galaxy network embedding: A hierarchical community structure
  preserving approach.
\newblock In {\em IJCAI}, pages 2079--2085, 2018.

\bibitem{keikha2018community}
Mohammad~Mehdi Keikha, Maseud Rahgozar, and Masoud Asadpour.
\newblock Community aware random walk for network embedding.
\newblock {\em Knowledge-Based Systems}, 148:47--54, 2018.

\bibitem{li2019learning}
Yu~Li, Ying Wang, Tingting Zhang, Jiawei Zhang, and Yi~Chang.
\newblock Learning network embedding with community structural information.
\newblock In {\em IJCAI}, pages 2937--2943, 2019.

\bibitem{ou2016asymmetric}
Mingdong Ou, Peng Cui, Jian Pei, Ziwei Zhang, and Wenwu Zhu.
\newblock Asymmetric transitivity preserving graph embedding.
\newblock In {\em Proceedings of the 22nd ACM SIGKDD international conference
  on Knowledge discovery and data mining}, pages 1105--1114, 2016.

\bibitem{ou2015non}
Mingdong Ou, Peng Cui, Fei Wang, Jun Wang, and Wenwu Zhu.
\newblock Non-transitive hashing with latent similarity components.
\newblock In {\em Proceedings of the 21th ACM SIGKDD International Conference
  on Knowledge Discovery and Data Mining}, pages 895--904, 2015.

\bibitem{lyu2017enhancing}
Tianshu Lyu, Yuan Zhang, and Yan Zhang.
\newblock Enhancing the network embedding quality with structural similarity.
\newblock In {\em Proceedings of the 2017 ACM on Conference on Information and
  Knowledge Management}, pages 147--156, 2017.

\bibitem{newman2006modularity}
Mark~EJ Newman.
\newblock Modularity and community structure in networks.
\newblock {\em Proceedings of the national academy of sciences},
  103(23):8577--8582, 2006.

\bibitem{blondel2008fast}
Vincent~D Blondel, Jean-Loup Guillaume, Renaud Lambiotte, and Etienne Lefebvre.
\newblock Fast unfolding of communities in large networks.
\newblock {\em Journal of statistical mechanics: theory and experiment},
  2008(10):P10008, 2008.

\bibitem{saxena2020survey}
Akrati Saxena.
\newblock A survey of evolving models for weighted complex networks based on
  their dynamics and evolution.
\newblock {\em arXiv preprint arXiv:2012.08166}, 2020.

\bibitem{onnela2007analysis}
Jukka-Pekka Onnela, Jari Saram{\"a}ki, J{\"o}rkki Hyv{\"o}nen, G{\'a}bor
  Szab{\'o}, M~Argollo De~Menezes, Kimmo Kaski, Albert-L{\'a}szl{\'o}
  Barab{\'a}si, and J{\'a}nos Kert{\'e}sz.
\newblock Analysis of a large-scale weighted network of one-to-one human
  communication.
\newblock {\em New Journal of Physics}, 9(6):179, 2007.

\bibitem{newman2001clustering}
Mark~EJ Newman.
\newblock Clustering and preferential attachment in growing networks.
\newblock {\em Physical review E}, 64(2):025102, 2001.

\bibitem{ravasz2002hierarchical}
Erzs{\'e}bet Ravasz, Anna~Lisa Somera, Dale~A Mongru, Zolt{\'a}n~N Oltvai, and
  A-L Barab{\'a}si.
\newblock Hierarchical organization of modularity in metabolic networks.
\newblock {\em science}, 297(5586):1551--1555, 2002.

\bibitem{lovasz1993random}
L{\'a}szl{\'o} Lov{\'a}sz et~al.
\newblock Random walks on graphs: A survey.
\newblock {\em Combinatorics, Paul erdos is eighty}, 2(1):1--46, 1993.

\bibitem{de2018combining}
Sam De~Winter, Tim Decuypere, Sandra Mitrovi{\'c}, Bart Baesens, and Jochen
  De~Weerdt.
\newblock Combining temporal aspects of dynamic networks with node2vec for a
  more efficient dynamic link prediction.
\newblock In {\em 2018 IEEE/ACM International Conference on Advances in Social
  Networks Analysis and Mining (ASONAM)}, pages 1234--1241. IEEE, 2018.

\bibitem{mikolov2013efficient}
Tomas Mikolov, Kai Chen, Greg Corrado, and Jeffrey Dean.
\newblock Efficient estimation of word representations in vector space.
\newblock {\em arXiv preprint arXiv:1301.3781}, 2013.

\bibitem{mikolov2013distributed}
Tomas Mikolov, Ilya Sutskever, Kai Chen, Greg Corrado, and Jeffrey Dean.
\newblock Distributed representations of words and phrases and their
  compositionality.
\newblock {\em NIPS}, 2013.

\bibitem{epasto2019single}
Alessandro Epasto and Bryan Perozzi.
\newblock Is a single embedding enough? learning node representations that
  capture multiple social contexts.
\newblock In {\em The World Wide Web Conference}, pages 394--404, 2019.

\bibitem{leskovec2012learning}
Jure Leskovec and Julian~J Mcauley.
\newblock Learning to discover social circles in ego networks.
\newblock In {\em Advances in neural information processing systems}, pages
  539--547, 2012.

\bibitem{leskovec2007graph}
Jure Leskovec, Jon Kleinberg, and Christos Faloutsos.
\newblock Graph evolution: Densification and shrinking diameters.
\newblock {\em ACM transactions on Knowledge Discovery from Data (TKDD)},
  1(1):2--es, 2007.

\bibitem{saxena2015understanding}
Akrati Saxena, SRS Iyengar, and Yayati Gupta.
\newblock Understanding spreading patterns on social networks based on network
  topology.
\newblock In {\em Proceedings of the 2015 IEEE/ACM International Conference on
  Advances in Social Networks Analysis and Mining 2015}, pages 1616--1617,
  2015.

\bibitem{gupta2016modeling}
Yayati Gupta, Akrati Saxena, Debarati Das, and SRS Iyengar.
\newblock Modeling memetics using edge diversity.
\newblock In {\em Complex Networks VII}, pages 187--198. Springer, 2016.

\bibitem{raghavan2007near}
Usha~Nandini Raghavan, R{\'e}ka Albert, and Soundar Kumara.
\newblock Near linear time algorithm to detect community structures in
  large-scale networks.
\newblock {\em Physical review E}, 76(3):036106, 2007.

\bibitem{cordasco2010community}
Gennaro Cordasco and Luisa Gargano.
\newblock Community detection via semi-synchronous label propagation
  algorithms.
\newblock In {\em 2010 IEEE International Workshop on: Business Applications of
  Social Network Analysis (BASNA)}, pages 1--8. IEEE, 2010.

\bibitem{pares2017fluid}
Ferran Par{\'e}s, Dario~Garcia Gasulla, Armand Vilalta, Jonatan Moreno, Eduard
  Ayguad{\'e}, Jes{\'u}s Labarta, Ulises Cort{\'e}s, and Toyotaro Suzumura.
\newblock Fluid communities: A competitive, scalable and diverse community
  detection algorithm.
\newblock In {\em International Conference on Complex Networks and their
  Applications}, pages 229--240. Springer, 2017.

\bibitem{zachary1977information}
Wayne~W Zachary.
\newblock An information flow model for conflict and fission in small groups.
\newblock {\em Journal of anthropological research}, pages 452--473, 1977.

\end{thebibliography}

\end{document}